\documentclass[preprint,aps,nofootinbib]{revtex4-1}
\pdfoutput=1
\usepackage{graphicx}

\usepackage{amsmath}
\usepackage{amsfonts}
\usepackage{amssymb}
\usepackage{color}
\usepackage{natbib}
\usepackage{hyperref}

\usepackage[utf8]{inputenc}

\def\be{\begin{equation}}
\def\ee{\end{equation}}

\usepackage{amsmath,amssymb}
\usepackage{slashed}
\usepackage{xcolor} 
\usepackage{graphicx}
\usepackage{url}

\def\lsim{\mathrel{\rlap{\lower4pt\hbox{\hskip1pt$\sim$}}
    \raise1pt\hbox{$<$}}}
\def\gsim{\mathrel{\rlap{\lower4pt\hbox{\hskip1pt$\sim$}}
    \raise1pt\hbox{$>$}}}

\newcommand{\beq}{\begin{equation}}
\newcommand{\eeq}{\end{equation}}
\newcommand{\bea}{\begin{eqnarray}}
\newcommand{\eea}{\end{eqnarray}}
\newcommand{\nn}{\nonumber \\ }

\graphicspath{{./Figures/}}

\newcommand{\fref}[1]{Fig.~\ref{fig:#1}} 
\newcommand{\eref}[1]{Eq.~\eqref{eq:#1}}

\newcommand{\tref}[1]{Table~\ref{tab:#1}}

\def\<{\langle}
\def\>{\rangle}

\def\bea{\begin{eqnarray}}
\def\eea{\end{eqnarray}}
\def\beq{\begin{equation}}
\def\eeq{\end{equation}}

\def\<{\langle}
\def\>{\rangle}

\def\ewd{{\mathcal{D}}}

\def\mO{\mathcal{O}}

\newcommand{\TeV}{\text{TeV}}

\newcommand{\Eq}[1]{Eq.~(\ref{#1})}

\begin{document}
\begin{flushright}
UG-FT 331/19,
CAFPE 201/19
\end{flushright}
\vspace*{.7cm}

\title{A Golden Probe of Nonlinear Higgs Dynamics}

\author{\vspace{.7cm} Da Liu$\,^{a}$, Ian Low$\,^{a,b}$, and Roberto Vega-Morales$\,^{c}$}
\affiliation{\vspace{.7cm}
\mbox{$\,^{a}$High Energy Physics Division, Argonne National Laboratory, Lemont, IL 60439, USA} \\
\mbox{$\,^{b}$Department of Physics and Astronomy, Northwestern University, Evanston, IL 60208, USA} \\
\mbox{$\,^{c}$CAFPE and Departamento de F\'{i}sica Te\'{o}rica y del Cosmos,}\\
\mbox{Universidad de Granada, Campus de Fuentenueva, E-18071 Granada, Spain} \\
 \vspace{0.3cm}
}

{
\begin{abstract}\vspace{0.3cm}
The most salient generic feature of a composite Higgs boson resides in the nonlinearity of its dynamics, which arises from degenerate vacua associated with the pseudo-Nambu-Goldstone (PNGB) nature of the Higgs boson. It has been shown that the nonlinear Higgs dynamics is universal in the IR and  controlled only by a single parameter $f$, the decay constant of the PNGB Higgs. In this work we perform a fit, for the first time, to Wilson coefficients of ${\cal O}(p^4)$ operators in the nonlinear  Lagrangian using the golden H $\to$ 4L decay channel. By utilizing both the ``rate" information in the signal strength and the "shape" information in the fully differential spectra, we  provide  limits on the Goldstone decay constant $f$, as well as ${\cal O}(p^4)$ Wilson coefficients, using Run 2 data at the LHC. In rate measurements alone, the golden channel prefers a negative $\xi =v^2/f^2$ corresponding to a non-compact coset structure. Including the shape information, we identify regions of parameter space where current LHC constraint on $f$ is still weak, allowing for $\xi \lesssim 0.5$ or $\xi \gtrsim -0.5$. We also comment on  future sensitivity at the high-luminosity upgrade of the LHC which could allow for simultaneous fits to multiple Wilson coefficients.

\end{abstract}
}

\maketitle
\tableofcontents

\section{Introduction}
\label{sec:intro}

The Higgs boson plays a central role in our understanding of physics at the electroweak scale, the energy scale being probed by the Large Hadron Collider (LHC). Precise measurement of its property is among the top priorities of  experimental programs at the LHC as well as any possible future electron-positron and  hadron colliders. A major goal of these efforts is to understand the microscopic nature of the 125 GeV Higgs boson:
\begin{itemize}
\item Is the Higgs boson a fundamental particle like the electron or a composite particle such as the pion?
\end{itemize}
Such a question could carry far reaching implications in our understanding of the Universe at the most fundamental scale, especially given the fact that no other fundamental scalar particles have been observed in nature. While current data at the LHC is consistent with the expectation of a Standard Model (SM) Higgs boson, the experimental uncertainty is still sizeable, on the order of 10 - 20\% or more \cite{Aad:2015gba,Aad:2019mbh,Sirunyan:2018koj}. Such a large uncertainty can hardly be characterized as ``precise" and the question of whether the 125 GeV Higgs boson is indeed {\em the} SM Higgs remains open.

There is a long history in the idea of a composite Higgs boson, dating back to the classic papers \cite{Kaplan:1983fs,Kaplan:1983sm} several decades ago, where the Higgs boson arises as a pseudo-Nambu-Goldstone boson (PNGB) like the pion in low-energy QCD. Models with a PNGB Higgs were revived and refined much later, via the little Higgs theories \cite{ArkaniHamed:2001nc,ArkaniHamed:2002qx,ArkaniHamed:2002qy} and the holographic Higgs models \cite{Contino:2003ve,Agashe:2004rs}. By now they are collectively referred to as the composite Higgs models.

There are numerous composite Higgs models \cite{Bellazzini:2014yua} which differ in several aspects. For example, the choice of symmetry breaking pattern $G/H$ where $G$ is the broken group in the UV and $H$ is the unbroken group containing the electroweak $SU(2)_L\times U(1)_Y$ gauge groups. The Higgs arises as a PNGB when $G$ is spontaneously broken to $H$ at a scale $\Lambda \sim 10$ TeV. One other aspect that varies greatly is the implementation of additional fermions associated with the third generation quarks in the SM, which  are introduced to reduce the UV sensitivity in the Higgs mass originating from the SM top quark. In some cases the new fermions do not even have to carry SM color charge \cite{Chacko:2005pe} or electroweak quantum numbers \cite{Cheng:2018gvu,Cohen:2018mgv}.

In spite of all these variations in model-building, there is one salient prediction that is generic to the entire class of composite Higgs models: 
\begin{itemize}
\item Nonlinear dynamics in Higgs  interactions with the electroweak gauge boson, as well as self-interactions carrying derivatives.
\end{itemize}
The origin of nonlinear Higgs interactions goes to the essence of a composite Higgs: the PNGB nature of the Higgs boson. It is the same nonlinear dynamics appearing in interactions of pions in chiral symmetry breaking, or any other Nambu-Goldstone bosons observed in nature. The theoretical tool employed to construct effective Lagrangians of Nambu-Goldstone bosons was developed half a century ago, in the seminal papers by Callan, Coleman, Wess and Zumino (CCWZ) \cite{Coleman:1969sm,Callan:1969sn}. In the CCWZ approach each symmetry breaking pattern $G/H$ results in a seemingly different effective Lagrangian, each with its own set of nonlinear interactions and experimental predictions.

Only in recent years was it realized that the nonlinear interaction of a PNGB has very little to do with the details of the UV group $G$ that is being spontaneously broken. This can be seen  either by imposing shift symmetries on the Nambu-Goldstone bosons in the IR \cite{Low:2014nga,Low:2014oga}, or by soft bootstrapping tree-level amplitudes of nonlinear sigma model (nl$\sigma$m) \cite{Cheung:2014dqa,Cheung:2016drk}. More specifically, the nonlinear interaction of Nambu-Goldstone bosons owes its presence to the existence of degenerate vacua in the deep IR and knows very little about the details of the broken Group $G$ \cite{Low:2018acv}. The only information on $G$ resides in the overall normalization of the Goldstone decay constant $f$, which can be taken as an input parameter in the low-energy.

Following this progress, the complete list of modifications to the couplings of one Higgs boson to two electroweak gauge bosons (HVV), two Higgs bosons with two electroweak gauge bosons (HHVV), one Higgs coupled to three electroweak gauge bosons (HVVV), as well as triple gauge boson couplings (TGC), were studied in Refs.~\cite{Liu:2018vel,Liu:2018qtb} up to four-derivative order and all orders in $1/f$. These corrections are the universal predictions of a composite Higgs boson. In particular, a set of ``universal relations" among the couplings were proposed which depend on only one input parameter $f$. Experimental confirmation of these universal relations would be a striking signal of the PNGB nature of the 125 GeV Higgs.

In this work we continue with the exploration of universal Higgs nonlinearity and its implications in Higgs coupling measurements. In particular, we will study the possibility of measuring and constraining Wilson coefficients of four-derivative operators in the effective Lagrangian of a composite Higgs boson using the H $\to$ 4L decay channel, the so-called ``Golden channel" because it is the clearest and cleanest among all Higgs decay channels. The small background contamination combined with the availability of full kinematic distributions of decay products offers not only a powerful probe of the spin and CP property of the Higgs boson \cite{Choi:2002jk,Buszello:2002uu}, but also a unique opportunity to employ advanced multivariate techniques \cite{Gainer:2011xz,Stolarski:2012ps,Chen:2012jy,Bolognesi:2012mm,Chen:2013ejz,Chen:2014pia,Chen:2014gka,Gainer:2014hha,Chen:2015rha,Chen:2015iha,Chen:2016ofc,Vega-Morales:2017pmm} to enhance the experimental sensitivity
\footnote{Analyses of Higgs couplings based on information geometry have also been shown to be powerful \cite{Brehmer:2016nyr,Brehmer:2017lrt}.}. We will use the publicly available LHC analyses in the 4L channel to probe and constrain, for the first time, Higgs nonlinear dynamics and  the associated Wilson coefficients. Furthermore, we demonstrate that the conventional approach of bounding the decay constant $f$ using the signal strength (total rate) in HVV measurements is incomplete, especially when effects of ${\cal O}(p^4)$ operators are included. 

This work is organized as the follows. In Section \ref{sec:nonlinear} we briefly review the universal Higgs nonlinearity, listing all operators modifying the HVV couplings up to ${\cal O}(p^4)$. We also map out the correspondence to the tensor structure basis used in the LHC analyses. Then in Section \ref{sec:expconst} we study the experimental constraints from both the rate and shape measurements in the H$\to$ 4L channel, as well as in H$\to$ Z$\gamma$ two body decays. Future sensitivity projections at the high-luminosity (HL) LHC are also provided in this section as well as a brief study of precision electroweak constraints. Finally we conclude in Section \ref{sec:conc}.

\section{Universal Higgs Nonlinearity}
\label{sec:nonlinear}

The effective Lagrangian of a PNGB Higgs boson contains two expansion parameters,
\begin{itemize}
\item Nonlinear expansion characterized by  ${\pi}/{f}$, where  $\pi$ denotes a generic Nambu-Goldstone field like a composite Higgs boson. The expansion in $\pi/f$ is highly nonlinear for a PNGB Higgs boson.
\item  Derivative expansion chararcterized by ${\partial_\mu}/{\Lambda}$. If some of the unbroken symmetry is gauged, one replaces the ordinary derivative by the gauge covariant derivative, $\partial_\mu \to D_\mu = \partial_\mu - i A_\mu$. In this power counting the gauge field $A_\mu$ then counts as one derivative.
\end{itemize}
This is similar to the chiral Lagrangian in low-energy QCD, which describes interactions of pions as PNGB's arising from spontaneously broken $SU(2)_L\times SU(2)_R$ chiral symmetry \cite{Georgi:1985kw}. In the chiral Lagrangian $f=f_\pi\sim 93$ MeV is the pion decay constant and $\Lambda = \Lambda_{\chi SB} \sim 1$ GeV is the scale where QCD becomes strongly coupled and chiral symmetry is spontaneously broken. In naive dimensional analysis (NDA) \cite{Manohar:1983md}, requiring quantum corrections from higher loops to be comparable in size to the derivative expansion in $\partial/\Lambda$ leads to the relation
\be
\Lambda \sim 4\pi f \ ,
\ee
which is saturated in a strongly interacting  theories with no relevant small adjustable parameters. Then the effective Lagrangian of Nambu-Goldstone bosons based on a nonlinear sigma model (nl$\sigma$m) is organized as follows
\be
\label{eq:ccwzeff}
S_{nl\sigma m} = \int d^4 x \, \Lambda^2 f^2 \, {\cal L}\left(\frac{\pi}{f}, \frac{\partial}{\Lambda}\right) =  \int d^4x\, \left[ {\cal L}^{(2)}+ {\cal L}^{(4)} +\cdots \right]\ ,
\ee
where ${\cal L}^{(n)}$  represent operators containing $n$ derivatives. In this work we focus on $n\le 4$. While ${\cal L}^{(n)}$ incorporates the derivative expansion at a well-defined order, the nonlinear expansion is resummed to all orders in $\pi/f$ in ${\cal L}^{(n)}$. Conventional wisdom from the CCWZ approach \cite{Coleman:1969sm,Callan:1969sn} has it that the nonlinear expansion realizes the spontaneously broken symmetries in the UV. The summation of operators to all orders in $\pi/f$ and only at a certain order in $\partial/\Lambda$ stems from requiring the resummed operators to have well-defined transformation properties under the broken symmetry. The nonlinear Higgs dynamics then follows. However, the CCWZ perspective obscures the universality in the Nambu-Goldstone interactions. It turns out that these nonlinear interactions  arise entirely from the presence of degenerate vacua in the IR and are insensitive to the coset structure $G/H$ in the UV \cite{Low:2014nga,Low:2014oga,Low:2018acv,Liu:2018vel}. The only parameter dependent on the coset structure is the normalization of $f$. In the end, by probing and measuring the Higgs nonlinear dynamics, one could potentially gain insight on a broad range of composite Higgs models, regardless of the $G/H$ coset.

Invoking the custodial invariance and focusing on the scenario where the 125 GeV Higgs transforms as a fundamental representation of an unbroken $SO(4)$ group, the leading two-derivative Lagrangian is  simple, upon gauging an $SU(2)_L\times U(1)_Y$ subgroup of $SO(4)$ \cite{Liu:2018vel,Liu:2018qtb},
\begin{align}
\label{eq:L2unitary}
{\cal L}^{(2)} &= \frac{1}{2} \partial_\mu h \partial^\mu h+ \frac{g^2f^2}{4} \sin^2( \theta + h/f)    \left(W^+_\mu W^{- \mu} + \frac{1}{2\cos^2 \theta_W} Z_\mu Z^\mu \right)\ ,
\end{align}
where $h$ is the 125 GeV Higgs boson. In the above $\sin\theta \equiv v/f$, where $v=246$ GeV, is the vacuum  misalignment angle and $W^\pm$ and $Z$ are the electroweak massive gauge bosons, whose masses can be read off from Eq.~(\ref{eq:L2unitary}):
\be
m_W =\frac{m_Z}{\cos\theta_W}=\frac12 g v = \frac12 g f \sin\theta\ .
\ee
As emphasized previously, ${\cal L}^{(2)}$ contains  two-derivative operators that are to all orders in $h/f$. In particular, coefficients of operators at different orders in $h/f$ are all predetermined by a single parameter $f$ and resummed to the form $\sin(\theta+h/f)$. Although not transparent at all, this particular form is enforced by four "shift symmetries" acting on the four components of the scalar $SU(2)_L$ doublet containing the neutral scalar $h$ \cite{Low:2014oga}.

Expanding Eq.~(\ref{eq:L2unitary}) in $h/v$, one obtains corrections to the SM HVV and HHVV couplings at the two-derivative level,
\be
\label{eq:L2exp}
\frac{1}{2} \partial_\mu h \partial^\mu h+ \left[2\sqrt{1-\xi} \, \frac{h}{v} + (1-2\xi )\,\frac{h^2}{v^2}+\cdots\right]\left( m_W^2  W^+_\mu W^{-\mu} + \frac{1}{2} m_Z^2 Z_\mu Z^{\mu} \right) \ .
\ee
where $\xi\equiv v^2/f^2$. Given that $f$ is an input parameter whose normalization is UV dependent, Eq.~(\ref{eq:L2exp}) is the basis for extracting $f$ using the signal strength in HVV coupling measurements.
\begin{figure}[t]
\begin{center}
\includegraphics[angle=270,width=10cm]{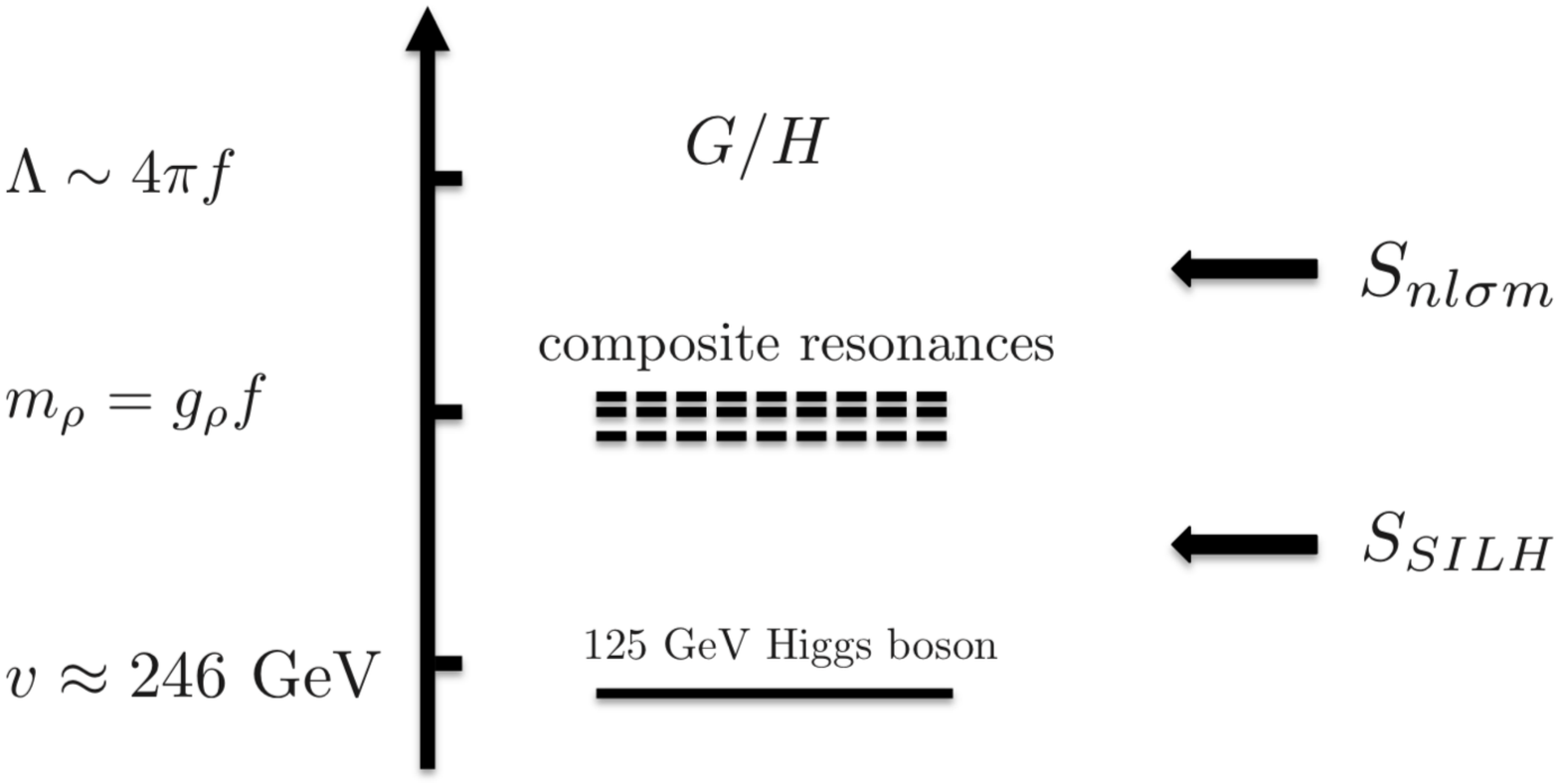}
\caption{\em Typical spectrum of composite Higgs models. The nl$\sigma$m Lagrangian is valid below $\Lambda\sim 4\pi f$, which is the highest cutoff in the model. Below the composite resonances the SILH effective Lagrangian is valid, although with identical nonlinear structure to the nl$\sigma$m Lagrangian, assuming a sufficient mass gap between $M_\rho$ and $f$.}
\label{fig:spectrum}
\end{center}
\end{figure}

In a strongly coupled theory there are typically composite resonances below the cutoff scale $\Lambda$, as is evident in low-energy QCD. In composite Higgs models these resonances are represented by the scale $M_\rho = g_\rho f$, where $1 \lesssim g_\rho \lesssim 4\pi$. Assuming there is a mass gap between $M_\rho$ and $f$, one could further integrate out the composite resonances and the resulting effective Lagrangian, which is valid at energy $E\sim f$, inherits the nonlinear interactions from above the scale $M_\rho$ \cite{Giudice:2007fh}. This effective theory is often referred to as the SILH Lagrangian,
\be
S_{SILH} = \int d^4 x \ M_\rho^2 f^2 \ {\cal L}\left(\frac{\pi}{f}, \frac{\partial}{M_\rho}\right) \ ,
\ee
which has the same nonlinear structure as in Eq.~(\ref{eq:ccwzeff}), but with $\Lambda \to M_\rho$. In particular, at the two-derivative level, Eq.~(\ref{eq:L2unitary}) remains unchanged even after integrating out $M_\rho$, and corrections to HVV and HHVV couplings are also unchanged. There are effects that could potentially spoil the nonlinear structure of the effective Lagrangian. These are related to ``explicit" symmetry breaking effects and examples include the Higgs potential and the Higgs coupling to fermions. However, as argued in Ref.~\cite{Liu:2018vel}, they would modify the nonlinearity HVV and HHVV couplings only at the loop-level.\footnote{These loop effects are non-universal and beyond the scope of current work.} The typical spectrum in a composite Higgs model is displayed in Fig.~\ref{fig:spectrum}.

\subsection{${\cal O}(p^4)$ Operators}
\label{subsec:powercount}

The two-derivative nonlinear Lagrangian in Eq.~(\ref{eq:L2unitary}) is valid when 
\be
\frac{v}{f} \lesssim 1 \quad \text{and} \quad \frac{E}{\Lambda} \ll 1 \ ,
\ee
where $E$ is the typical energy scale probed by the experiment. In particular, resumming contributions to all orders in $v/f$ allows for an  $f$ not too far above the weak scale set by $v\approx 246$ GeV. We can further extend the range of validity of the effective action by including higher order terms in the derivative expansion, such as ${\cal O}(p^4)$ operators. Chiral Lagrangian operators that are ${\cal O}(p^4)$ and to all orders in $\pi/f$ have been enumerated in Ref.~\cite{Gasser:1983yg}. In the context of composite Higgs bosons the relevant four-derivative operators for Higgs couplings were given in Refs.~\cite{Liu:2018vel,Liu:2018qtb} in the unitary gauge.

More specifically, the SILH Lagrangian at ${\cal O}(p^4)$, including the full nonlinearity structure, can be written as
\be
\label{eq:lag4}
S_{SILH}^{(4)}= \int d^4x \ M_\rho^2\, f^2\ {\cal L}^{(4)}\left(\frac{\pi}f, \frac{D}{M_\rho}\right)=\int d^4x\ \sum_i \frac{c_i}{g_\rho^2} O_i \ ,
\ee 
where $c_i$ are expected to be order unity constants parameterizing the incalculable UV physics at the scale $\Lambda \sim 4\pi f$. In some cases operators  contributing to  couplings of neutral particles and an on-shell photon are further suppressed by additional loop factors. In total there are 7 operators labelled by $O_1, O_2, O_3, O_4^\pm$ and $O_5^\pm$, each with the corresponding unknown Wilson coefficients $c_i$. These 7 Wilson coefficients contribute to a dozen different observables that can be measured in HVV and HHVV couplings \cite{Liu:2018vel,Liu:2018qtb}. Focusing on  those relevant for HVV couplings we have, in the unitary gauge,
\bea
\label{eq:efthvv}
{\cal L}^{(1h)} &=&  \frac{m_W^2}{M_\rho^2} \left[ C_1^h\   \frac{h}{v} Z_{\mu} \ewd^{\mu \nu} Z_{\nu} + C_2^h\  \frac{h}{v}  Z_{\mu\nu} Z^{\mu\nu} + C_3^h\ \frac{h}{v}  Z_{\mu} \ewd^{\mu \nu} A_{\nu} \right.\nonumber\\
&&\qquad \left.+ C_4^h \ \frac{h}{v}  Z_{\mu\nu} A^{\mu\nu} + C_5^h\ \frac{h}{v} ( W^+_{\mu} \ewd^{\mu \nu} W^-_{\nu} + \text{h.c.})
+ C_6^h\  \frac{h}{v}  W^+_{\mu\nu} W^{-\mu\nu} \right] \ ,
\eea
where $\ewd =\partial^\mu \partial^\nu - \eta^{\mu\nu} \partial^2$. Although there are six coefficients $C_i^h, i=1,\cdots, 6$  in the unitary gauge, they are secretly related by only five Wilson coefficients residing in Eq.~(\ref{eq:lag4}), as well as the input parameter defined by $\sin\theta=\sqrt{\xi}=v/f$, in a composite Higgs model:
\bea
C_1^h &=& \frac{4 c_{ 2 w} }{c^2_{w}}  \left(-2 c_3  + c_4^- \right)  +\frac{4}{c^2_{w}}   c_4^+\cos \theta \ , \\
C_2^h&=& -\frac{2  c_{2 w}}{c^2_{w}}  \left( c_4^-  + 2 c_5^- \right) -\frac{2 }{c^2_{w}}   \left( c_4^+  - 2 c_5^+ \right) \cos \theta  \ , \\
C_3^h &=& 8  \left( - 2 c_3  +   c_4^- \right)  t_{ w} \ ,\\
C_4^h&=& -4  \left(  c_4^-   +2 c_5^- \right)  t_{ w} \ , \\
C_5^h &=&4 (-2 c_3  + c_4^- ) +  4 c_4^+ \cos \theta \ , \\
C_6^h &=&  -4( c_4^- + 2 c_5^-)   -4 \left( c_4^+  - 2 c_5^+\right) \cos \theta \ .
\eea
In the above  $c_w, c_{2w}, t_w$ denote $\cos\theta_W, \cos2\theta_W, \tan\theta_W$ respectively, where $\theta_W$ is the weak mixing angle.

\subsection{H$\to$ 4L Tensor Structure}
\label{subsec:hvv}
 
 The operators listed in Eq.~(\ref{eq:efthvv}) enter into H$\to$ 4L decays and manifest themselves through kinematic distributions of the decay product. For our analysis we utilize the ``Golden Channel'' analysis framework developed in Refs.~\cite{Gainer:2011xz,Chen:2012jy,Chen:2013ejz,Chen:2014pia,Chen:2014gka,Chen:2015rha,Chen:2015iha}  which parametrizes the effective Higgs boson couplings to pairs of neutral vector boson pairs in terms of the Lorentz tensor structures, 
\be
\label{eq:vert}
 \Gamma_{V}^{\mu \nu}  =   {1 \over v} \left [  
A_1^{V} m_Z^2 g^{\mu\nu}  +  A_2^{V} (k_1^\nu k_2^\mu - k_1\cdot k_2 g^{\mu\nu})
+ A_3^{V} \epsilon^{\mu\nu\alpha\beta} k_{1\alpha} k_{2\beta}
+  (A_4^{V} k_1^2  + \bar{A}_4^{V} k_2^2 )g^{\mu \nu}   \right ]\ ,
\ee
where $V = (ZZ, Z\gamma, \gamma Z, \gamma\gamma)$. We assume massless leptons, using the notation and conventions defined in~\cite{Chen:2014gka}, but have also included the $A_4^V$ tensor structures which were not included previously. Note that electromagnetic gauge invariance requires $A_1^{\gamma\gamma} = A_1^{Z\gamma} = A_4^{\gamma\gamma} = 0$. In general the $A_i^V$ can be momentum dependent form factors which are functions of Lorentz invariant products of the external momenta. For our purpose it is sufficient to take them to be real and constant coefficients in which case we have $A_{4} = \bar{A}_{4}$. Furthermore, in the massless lepton case we have $Z^\mu \partial^\nu V_{\mu\nu} = Z_\mu \mathcal{D}^{\mu\nu} V_{\nu}$. Then the relation between $A_i^V$ and the Wilson coefficients defined in Eq.~(\ref{eq:efthvv}) is quite simple:
\bea\label{eq:AtoNLH}
A_1^{ZZ} &=&  2 \sqrt{1-\xi}  , \nn
A_2^{ZZ} &=& 4 \frac{m_W^2}{M_\rho^2} C_2^h, \nn
A_4^{ZZ} &=& \frac{m_W^2}{M_\rho^2} C_1^h , \\
A_2^{Z\gamma} &=& 4 \frac{m_W^2}{M_\rho^2} C_4^h ,\nn
A_4^{Z\gamma}  &=& \frac{m_W^2}{M_\rho^2} C_3^h .\nonumber
\eea
For simplicity we have kept only CP-even terms and neglected effective couplings to pairs of photons in~\eref{vert}, which do not appear in universal Higgs nonlinearity, though it would be straightforward to include them in the fit. Note that as $\xi \to 0$ we recover the SM value for the tree level HZZ coupling, $A_1^{ZZ} = 2$. The important observation is:
\begin{itemize}
\item Only $A_1^{ZZ}$ is directly related to $\xi$; the other tensor structures are dependent on both $M_\rho$ and $C_i^h$.
\end{itemize}
At the leading order in derivative expansion, the signal strength in HVV coupling measurements is attributed entirely to  $A_1^{ZZ}$, which is then used to constrain $\xi$ \cite{deBlas:2018tjm}. We see this is not the case anymore when ${\cal O}(p^4)$ effects are included. It turns out that all couplings in Eq.~(\ref{eq:AtoNLH}) can be extracted from the fully differential spectra of H$\to$ 4L decays. In what follows we use CMS H$\to$ 4L data to constrain the  $A_i^V$ coefficients, which then translate into limits on $\xi$ and the Wilson coefficients $C_i^h$.

\section{Experimental Constraints}
\label{sec:expconst}

\subsection{The Golden Channel: H $\to$ 4L}
\label{subsec:hto4l}

In this subsection we will use both the ``rate information," which pertains to the signal strength measured in H$\to$ 4L channel, and the ``shape information," as contained in the differential spectra of final state leptons, to place constraints on the Wilson coefficients in the nonlinear Higgs Lagrangian in Eq.~(\ref{eq:efthvv}). In particular, we utilize the fully differential decay width analytically computed in~\cite{Chen:2012jy,Chen:2013ejz} for H $\to 2e2\mu, 4e$, and $4\mu$ assuming on-shell decay of the Higgs boson. Interference effects between the different tensor structures in~\eref{vert}, as well as among identical final states in the case of $4e$ and $4\mu$, have been fully accounted for. Before examining current CMS constraints, we briefly review the H$\to$ 4L partial decay width and fully differential spectra.

For our purpose it is convenient to single out the $A_i^{V}$ dependence in the differential decay width, which can be  written schematically as
\bea
\label{eq:diffwidth}
\frac{d\Gamma_{\text{H$\to$4L}}}{d\mathcal{O}} 
= \sum_{ij}  A_{i} A_j^{\ast} 
\times \frac{d\hat\Gamma_{ij}}{d\mathcal{O}}\ ,
\eea
where $\mathcal{O}$ represents all observables available in the H$\to$ 4L decay channel \cite{Chen:2012jy,Chen:2013ejz} and $i,j$ sum over all of the possible tensor structures  in Eq.~(\ref{eq:vert}). The advantage of doing so is then the remaining quantities ${d\hat\Gamma_{ij}}/{d\mathcal{O}}$ can be calculated and integrated over a particular phase space. We can then define the `sub-widths' for each combination of $A_i A_j^{\ast}$ as
\bea
\label{eq:totw}
\hat{\Gamma}_{ij} = \int \frac{d\hat\Gamma_{ij}}{d\mathcal{O}} d\mathcal{O} \ ,
\eea
which is just a numerical constant once a selection cut over the phase space is chosen. It is worth emphasizing that the sub-widths could be negative for certain combinations of tensor structures when they interfere destructively. However, the partial width written as the sum
\bea
\label{eq:norm}
\Gamma_{\text{H$\to$4L}}  = \sum_{ij}  A_{i} A_{j}^{\ast}
\times \hat{\Gamma}_{ij} ,
\eea
must be positive and is now a function of the effective couplings and the phase space cuts. \eref{diffwidth} and \eref{norm} allow for a full reconstruction of the differential decay spectra and the partial width of H$\to$ 4L, respectively.

It will be convenient to normalise $\Gamma_{\rm H\to4L}$ to the (tree level) SM expectation, which corresponds to $A_1^{ZZ} = 2$ and all other couplings set to zero,
\bea
\label{eq:normratio}
\mathcal{R}_{\text{4L}} \equiv \frac{\Gamma_{\text{H$\to$4L}}}{\Gamma^{\text{SM}}_{\rm H\rightarrow 4L} }
= \sum_{ij}  A_{i} A_j^{\ast}
\times  \frac{\hat{\Gamma}_{ij}}{\Gamma^{\text{SM}}_{\rm H\rightarrow 4L} } \ .
\eea
Performing the integration over phase space we obtain the normalized partial width in Eq.~(\ref{eq:normratio}) for the $2e2\mu$ and $4e/4\mu$ channels respectively,
\bea\label{eq:R4lAi}
\mathcal{R}_{2e2\mu} &=&
0.25 |A_1^{ZZ}|^2 
+ 0.00092 A_1^{ZZ} A_2^{\gamma\gamma} 
+ 2.97 |A_2^{\gamma\gamma}|^2 
+ 0.0868 A_1^{ZZ} A_2^{Z\gamma} 
- 0.1253 A_2^{\gamma\gamma} A_2^{Z\gamma} \nn
 &+& 9.30 |A_2^{Z\gamma}|^2 
 - 0.0696 A_1^{ZZ} A_2^{ZZ} 
 + 0.0000346 A_2^{\gamma\gamma} A_2^{ZZ}  
- 0.0292 A_2^{Z\gamma} A_2^{ZZ}  
+ 0.0253 |A_2^{ZZ}|^2 \nn
 &+& 2.508 |A_3^{\gamma\gamma}|^2 
 - 0.0958 A_3^{\gamma\gamma} A_3^{Z\gamma}  
 + 5.14 |A_3^{Z\gamma}|^2 
 +  0.0003935 A_3^{\gamma\gamma} A_3^{ZZ} 
 - 0.01301 A_3^{Z\gamma} A_3^{ZZ} \nn
 &+& 0.009033 |A_3^{ZZ}|^2 
 - 0.03134 A_1^{ZZ} A_4^{Z\gamma} 
 + 0.03914 A_2^{\gamma\gamma} A_4^{Z\gamma} 
 - 1.815 A_2^{Z\gamma} A_4^{Z\gamma}  \nn
 &+& 0.0007644 A_2^{\gamma\gamma} A_4^{ZZ} 
 + 0.008514 A_2^{ZZ} A_4^{Z\gamma} 
 + 0.6082 |A_4^{Z\gamma}|^2 
 + 0.2445 A_1^{ZZ} A_4^{ZZ}  \nn
 &+& 0.08802 A_2^{Z\gamma} A_4^{ZZ} 
 - 0.06939 A_2^{ZZ} A_4^{ZZ}  
 -  0.02968 A_4^{Z\gamma} A_4^{ZZ} 
 + 0.2468 |A_4^{ZZ}|^2 , \\
\mathcal{R}_{4e/4\mu} &=&
0.25 |A_1^{ZZ}|^2 
- 0.076 A_1^{ZZ} A_2^{\gamma\gamma} 
+ 15.19 |A_2^{\gamma\gamma}|^2 
+ 0.088 A_1^{ZZ} A_2^{Z\gamma} 
- 0.1845 A_2^{\gamma\gamma} A_2^{Z\gamma} \nn
 &+& 8.34 |A_2^{Z\gamma}|^2 
 - 0.0659 A_1^{ZZ} A_2^{ZZ} 
 - 0.00567 A_2^{\gamma\gamma} A_2^{ZZ}  
- 0.02475 A_2^{Z\gamma} A_2^{ZZ}  
+ 0.02168 |A_2^{ZZ}|^2 \nn
 &+& 14.14 |A_3^{\gamma\gamma}|^2 
 - 0.1297 A_3^{\gamma\gamma} A_3^{Z\gamma}  
 + 4.507 |A_3^{Z\gamma}|^2 
 -  0.0248 A_3^{\gamma\gamma} A_3^{ZZ} 
 - 0.00875 A_3^{Z\gamma} A_3^{ZZ} \nn
 &+& 0.00704 |A_3^{ZZ}|^2 
 - 0.0423 A_1^{ZZ} A_4^{Z\gamma} 
 + 0.05394 A_2^{\gamma\gamma} A_4^{Z\gamma} 
 - 1.74 A_2^{Z\gamma} A_4^{Z\gamma}  \nn
 &-& 0.0334 A_2^{\gamma\gamma} A_4^{ZZ} 
 + 0.0103 A_2^{ZZ} A_4^{Z\gamma} 
 + 0.6957 |A_4^{Z\gamma}|^2 
 + 0.2316 A_1^{ZZ} A_4^{ZZ}  \nn
 &+& 0.08167 A_2^{Z\gamma} A_4^{ZZ} 
 - 0.06262 A_2^{ZZ} A_4^{ZZ}  
 -  0.03494 A_4^{Z\gamma} A_4^{ZZ} 
 + 0.2251 |A_4^{ZZ}|^2 , \nonumber
\eea
with cuts and reconstruction corresponding to `CMS-like' phase space selections~\cite{Khachatryan:2014kca,Sirunyan:2019twz,Sirunyan:2017tqd}. In this work we have included the $A_4^{ZZ/Z\gamma}$ couplings, which were not included in previous studies~\cite{Chen:2014gka,Chen:2015iha}. For completeness we have included all of the operators in Eq.~(\ref{eq:vert}), though below we will focus on those relevant for the nonlinear Higgs dynamics we are interested in. Note terms linear in the CP-odd couplings (${A}_3^{VV}$) do not appear because they integrate to (nearly) zero when choosing CMS-like selection cuts, which reflects the fact that ``rate" measurements are not sensitive to CP violation. In this regard, one could employ the shape information in the differential spectra \cite{Chen:2014gka,Chen:2015rha,Chen:2015iha} or the construction of forward-backward asymmetries~\cite{Chen:2014ona} to probe CP violation, but we do not explore this possibility here.

In the CMS analyses in Refs.~\cite{Khachatryan:2014kca,Sirunyan:2019twz}, the signal strength in H$\to$ 4L is measured in two categories, depending on whether the production channels involve Higgs couplings to fermions (ggH and ttH channels) or Higgs couplings to electroweak bosons (VBF and VH channels),
\be
\mu_i^{\text{4L}} =
\frac{\sigma(i \to \text{H} \to \text{4L}) }{\sigma(i \to \text{H} \to \text{4L})_{\rm{SM}}} 
=
\frac{ \sigma (i\to \text{H}) }{\sigma (i\to \text{H})_{\rm{SM}}}
\times 
\frac{ \Gamma^{\rm{SM}}_{\text{H}} }{ \Gamma_{\text{H}}  }
\times 
\frac{\Gamma_{\text{H$\to$4L}}}{\Gamma^{\text{SM}}_{\rm H \rightarrow 4L} }\  ,
\ee
where $i = F, V$ represents the ggH+ttH channels and VBF+VH channels, respectively. In the following we will focus on the fermionic production channels, ggH+ttH, for two reasons: 1) ggH is by far the dominant production channel of the 125 GeV Higgs at the LHC and the uncertainty is smaller and 2) any modification in HVV couplings will enter into both the production and decay amplitudes in the VBF+VH channels. For simplicity we assume the production cross-sections in ggH+ttH channels are equal to their SM values as well as the total Higgs width. Under these assumptions the deviation in ``rate measurements" in $\mu_F^{\text{4L}}$ arises entirely from the decay amplitudes:
\bea\label{eq:Rimu4L}
\mu_F^{\text{4L}} = \frac{\Gamma_{\text{H$\to$4L}}}{\Gamma^{\text{SM}}_{\rm H \rightarrow 4L} } = \mathcal{R}_{\rm 4L} \ .
\eea

\begin{figure}[t]
\begin{center}
\includegraphics[width=8cm]{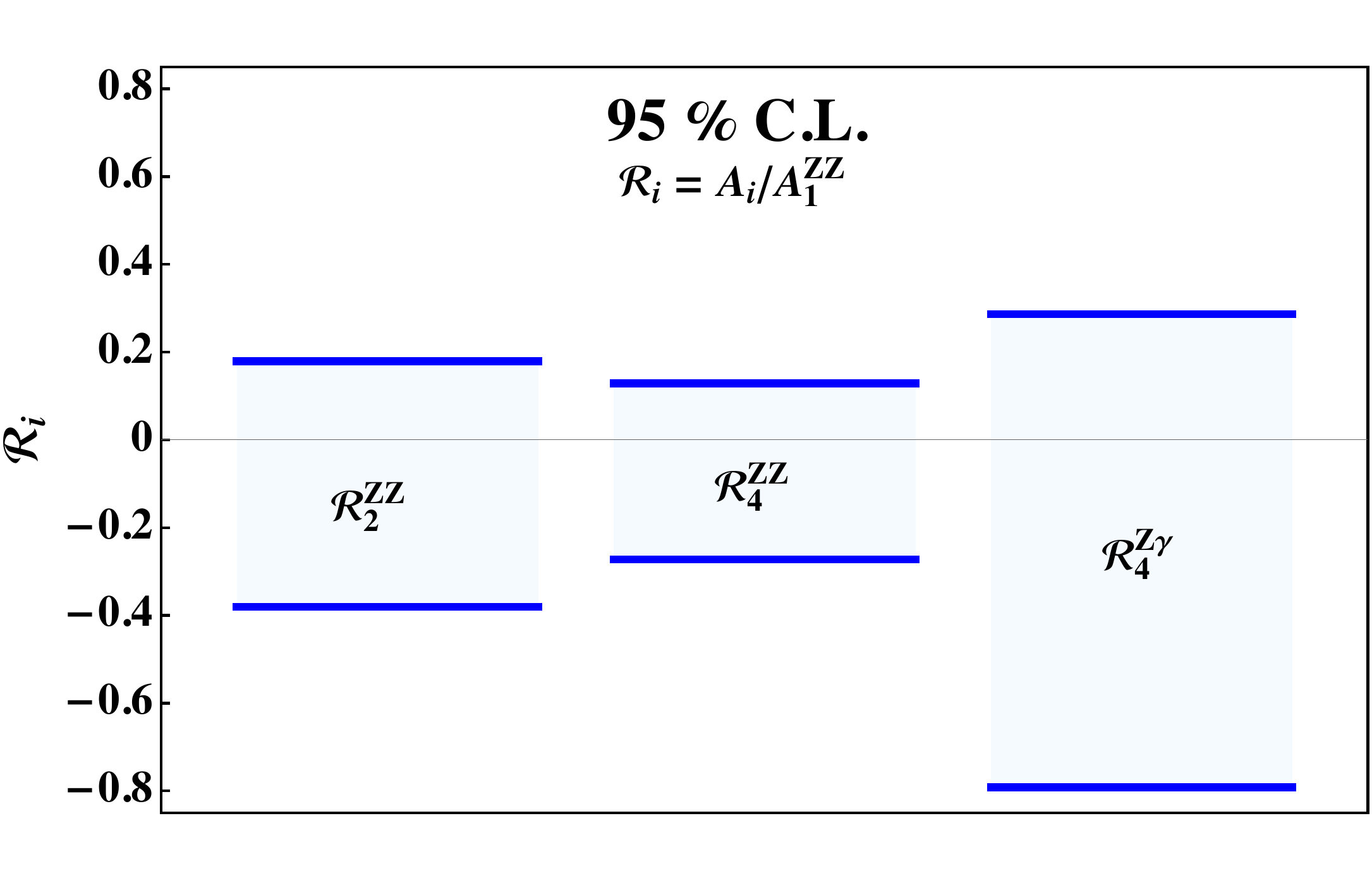}
\caption{\em Visual representation of bounds from Table \ref{tab:Abounds}.}
\label{fig:xifit0}
\end{center}
\end{figure}

For completeness we briefly summarize the procedures taken by the CMS collaboration in Refs.~\cite{Khachatryan:2014kca,Sirunyan:2019twz}, which we refer the reader to for details. For CP-even couplings, which we focus on in this work, CMS built three multi-variate likelihood functions, each optimized for a particular anomalous Higgs coupling in Eq.~(\ref{eq:vert}): $A_2^{ZZ},  A_4^{ZZ}$ and $A_4^{Z\gamma}$. In each likelihood function all  anomalous couplings, other than the one the likelihood function is specifically optimized for, are set to zero \cite{gritsan2019}. In other words, in these analyses the anomalous HVV couplings are turned on only one at a time. (See Refs.~\cite{Chen:2013ejz,Chen:2014pia} for a framework that allows for simultaneous measurements of several anomalous Higgs couplings at the same time.) The likelihood function allows one to constrain and fit: 1) the rate (signal strength) in H$\to$ 4L decays and 2) the ``fraction" of the observed 4L events originated from the anomalous HVV coupling. It turns out that the best fit value for the three CP-even anomalous HVV couplings analyzed in Refs.~\cite{Khachatryan:2014kca,Sirunyan:2019twz} all have central values extremely close to zero, albeit with varying degrees of uncertainties.  As a result, CMS provided four different fits to the signal strength in 4L channel, under the assumption of vanishing anomalous HVV couplings. The relevant CMS results are summarized in Table \ref{tab:mucms}, where we have computed the 95\% C.L. from the 67\% C.L. by assuming a Gaussian distribution.  As for the shape measurements, CMS provided limits on the ratio
\footnote{Note we use a slightly different normalization than in the CMS analysis~\cite{Khachatryan:2014kca,Sirunyan:2019twz}.}
\be
\label{eq:Aratio}
\mathcal{R}^V_i \equiv A^V_i/A_1^{ZZ} ,
\ee
from the differential spectra in the 4L decays. This is possible because different tensor structures in Eq.~(\ref{eq:vert}) result in different shapes in the differential distributions. We quote the results in Table \ref{tab:Abounds} and give a visual representation of the bounds in Fig.~\ref{fig:xifit0}.

In their Run 2 analyses CMS did not provide a fit to $A_2^{\gamma\gamma/Z\gamma}$ \cite{Khachatryan:2014kca,Sirunyan:2019twz}, because these are better constrained from direct H$\to\gamma\gamma$ and H$\to$Z$\gamma$ two body decays. It turns out that only $A_2^{Z\gamma}$ is universal in nonlinear Higgs dynamics, as $A_2^{\gamma\gamma}$ requires shift-symmetry breaking effects. In Section \ref{sect:Zgamma} we perform a separate fit to $A_2^{Z\gamma}$  using the direct Z$\gamma$ channel.

\begin{table}[t]
\begin{center}
\begin{tabular}{|c|c|c|}
\hline
$\mu_{F}^{\text{4L}}({A}_i)$  &~Best Fit at ${A}_i = 0$~&~$95\%$\,C.L.~\cite{Sirunyan:2019twz}~\\ \hline 
$\mu_{F}^{\text{4L}}({A}_2^{ZZ})$ & 1.19 & $[0.857, 1.602]$ \\ \hline
$\mu_{F}^{\text{4L}}({A}_{4}^{ZZ})$ & 1.26 & $[0.907, 1.652]$ \\ \hline
$\mu_{F}^{\text{4L}}({A}_{4}^{Z\gamma})$ & 1.24 & $[0.907, 1.612]$ \\ \hline
\end{tabular}
\caption{\em Summary of CMS rate measurements of H$\to$ 4L from ggh+ttH production channels \cite{Khachatryan:2014kca,Sirunyan:2019twz}. Each measurement is optimized with respect to a particular anomalous HVV coupling as indicated. However, the fit is performed assuming $A_i^V=0$.}
\label{tab:mucms}
\end{center}
\end{table}
\begin{table}[thb]
\begin{center}
\begin{tabular}{|c|c|c|}
\hline
$\mathcal{R}_i = A_i/A_1^{ZZ}$ & Constraint & $95\%$\,CL \\ \hline 
$\mathcal{R}_2^{ZZ}$ & $\Gamma_H =  \Gamma_H^{\rm{SM}}$ &~$[-0.381, 0.180]$~\cite{Sirunyan:2019twz} \\ \hline 
$\mathcal{R}_4^{ZZ}$ & $\Gamma_H =  \Gamma_H^{\rm{SM}}$ &~$[-0.272, 0.129]$~\cite{Sirunyan:2019twz} \\ \hline
$\mathcal{R}_4^{Z\gamma}$ & On-shell Events &~$[-0.792, 0.287]$~\cite{Sirunyan:2019twz} \\ \hline
\end{tabular}
\caption{\em Summary of CMS shape measurements on ratios of anomalous HVV couplings \cite{Khachatryan:2014kca,Sirunyan:2019twz}. In the middle column we indicate the assumptions on the Higgs total width. }
\label{tab:Abounds}
\end{center}
\end{table}
%

It is worth emphasising that, in order to properly constrain nonlinear Higgs dynamics, one would need to adapt the present CMS analyses. In particular, we would like to
\begin{enumerate}
\item fit the magnitude of the $A_i^V$ coefficient  in Eq.~(\ref{eq:vert}), instead of just the fraction.
\item allow all anomalous HVV couplings to be present at the same time.
\end{enumerate}
It would be of interest to study the impact on  the signal strength ($\mu_{F}$) measurement by relaxing the assumption of vanishing  anomalous HVV couplings. In what follows we will perform a limited analysis within the framework of the CMS analyses, and await a more comprehensive experimental study in the future.

\subsubsection{Constraints From Rate Measurements}\label{sec:fits1}

\begin{figure}[t]
\begin{center}
\includegraphics[width=8cm]{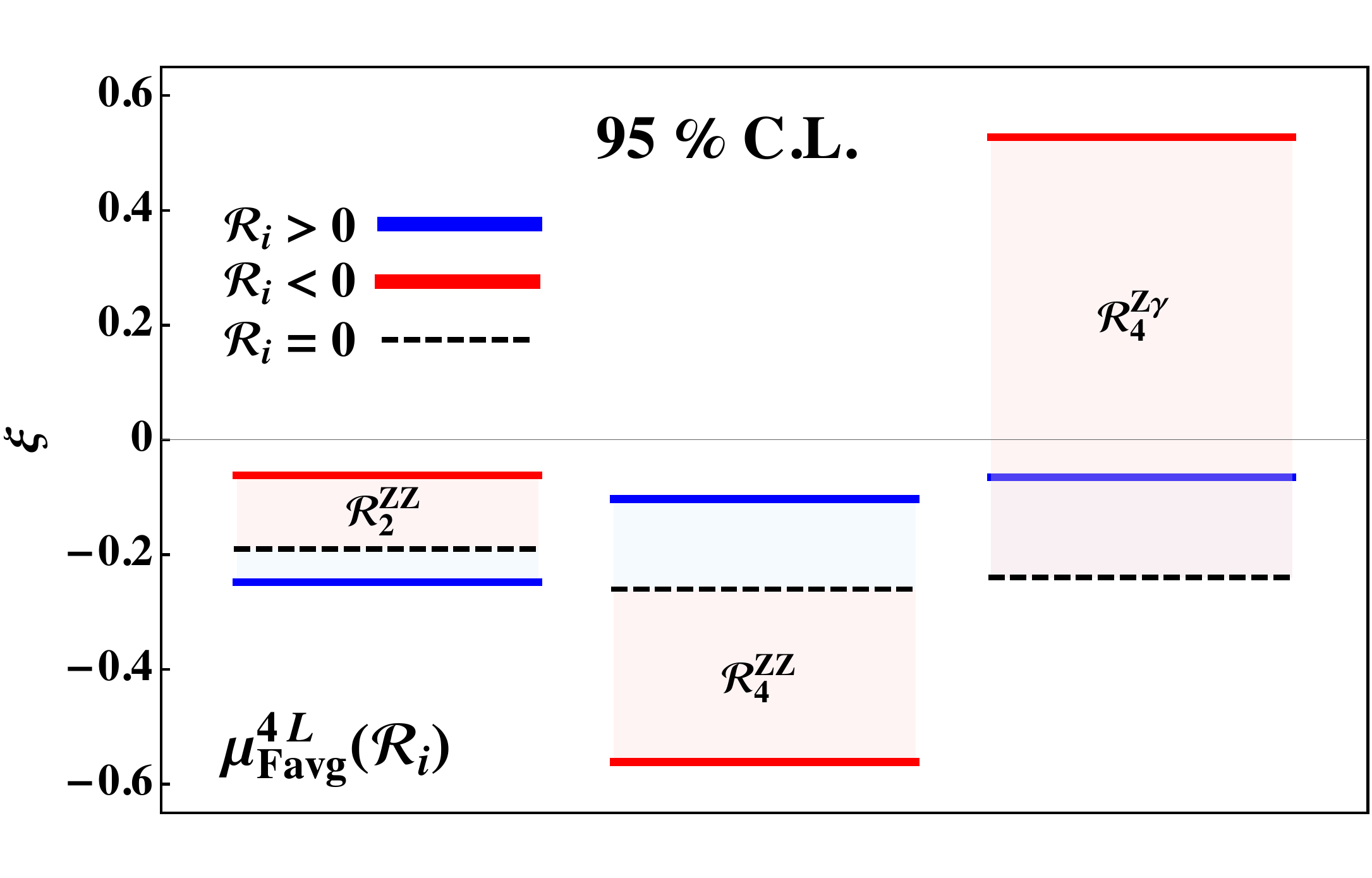}
\includegraphics[width=8cm]{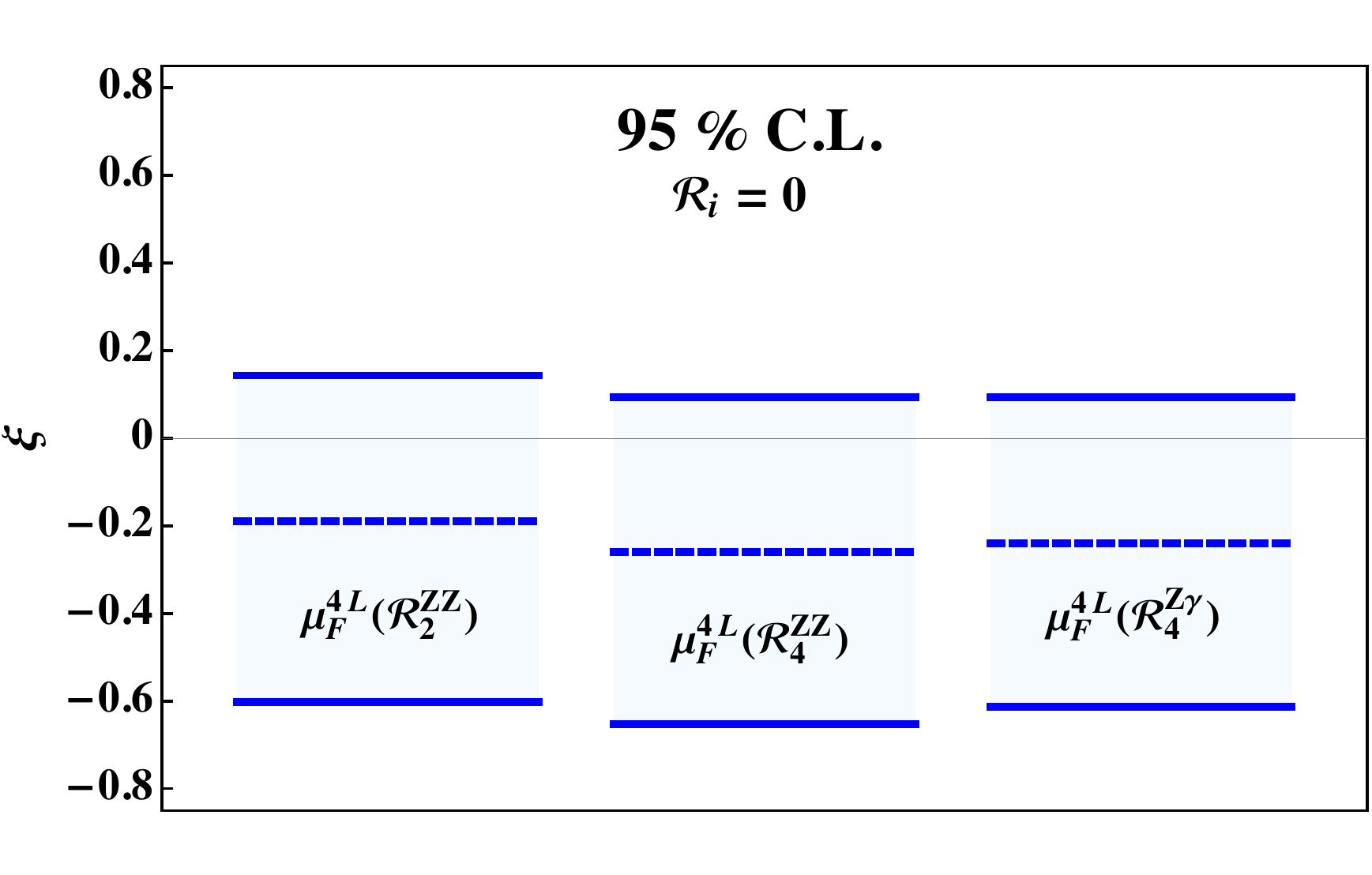}
\caption{\em {\bf Right:}~Fits to $\xi$ using the rate information in the H$\to$4L channel, assuming the anomalous HVV couplings vanish.~{\bf Left:} Fits to $\xi$ using the shape information and the constraints on ${\cal R}_i^V$. See the main text for more details.}
\label{fig:xifit}
\end{center}
\end{figure}

Using the CMS bounds given in~\tref{mucms} and~\tref{Abounds} we can now study limits derived from rate measurements. We restrict ourselves to the  CP-even HVV couplings $A_1^{ZZ}, A_2^{ZZ}, A_4^{ZZ}$ and $A_4^{Z\gamma}$, and  trade  $A_1^{ZZ}$ for the nonlinear parameter $\xi$. Then \eref{Rimu4L} can be written as
\bea\label{eq:R4lAishort}
\mu_F^{\text{4L}}(\mathcal{R}_i) &=&
4(1 - \xi)
\Big(0.25
- 0.0696 \mathcal{R}_2^{ZZ} 
+ 0.2445 \mathcal{R}_4^{ZZ} 
- 0.0313 \mathcal{R}_4^{Z\gamma} \nn
&&+ 0.0253 |\mathcal{R}_2^{ZZ}|^2 
+ 0.2468 |\mathcal{R}_4^{ZZ}|^2
+ 0.6082 |\mathcal{R}_4^{Z\gamma}|^2 \nn
&&- 0.06939 \mathcal{R}_2^{ZZ} \mathcal{R}_4^{ZZ} 
- 0.02968 \mathcal{R}_4^{Z\gamma} \mathcal{R}_4^{ZZ} 
+ 0.008514 \mathcal{R}_2^{ZZ} \mathcal{R}_4^{Z\gamma} 
\Big) \ ,
\eea
where we have used only the $2e2\mu$ normalized partial width in Eq.~(\ref{eq:R4lAi}) with similar results obtained using the $4e$ channel for the operators of interest in this study. With Eq.~(\ref{eq:R4lAishort}) we can perform two kinds of fits to $\xi$: 
\begin{enumerate}
\item In the right-panel of Fig.~\ref{fig:xifit}, we present bounds on $\xi$ by setting all ${\cal R}_i=0$ and applying the measured $\mu_F^{\text{4L}}$ in Table \ref{tab:mucms}, where the constraints are obtained assuming that the anomalous HVV coupling vanishes. The values of $\xi$ for the  different analysis vary because different categorization and observables are utilized. 
\item In the left-panel of Fig.~\ref{fig:xifit}, we use the central value of $\mu_F^{\text{4L}}$ from Table \ref{tab:mucms} in the left-hand side of Eq.~(\ref{eq:R4lAishort}) and plug in the experimental limits on the corresponding ${\cal R}_i$ from Table \ref{tab:Abounds} in the right-hand side. We then invert Eq.~(\ref{eq:R4lAishort}) to derive a limit on $\xi$.
\end{enumerate}
The moral of these two fits on $\xi$ is very different, and neither is perfect. 

In the first fit, we assume all the observed 4L events came from the $g^{\mu\nu}$ tensor structure in Eq.~(\ref{eq:vert}), which is similar in spirit to the conventional approach of using the signal strength to constrain $\xi$ without including ${\cal O}(p^4)$ effects. We have seen in Section \ref{sec:nonlinear} that this holds only at leading order in the derivative expansion. One could justify this approach somewhat by pointing out that the central values of the anomalous HVV couplings from the shape information are all extremely close to zero \cite{Sirunyan:2019twz}. However the uncertainties remain significant, as can be seen in Table  \ref{tab:Abounds}, and it is not clear how to interpret $\mu_F^{\text{4L}}$ when the anomalous couplings are turned on. 

In the second fit, we attempted to subtract out the 4L events originating from the anomalous HVV coupling, by using the experimental limit on ${\cal R}_i^V$ in Table  \ref{tab:Abounds}. The remaining events then can be interpreted as arising entirely from the $g^{\mu\nu}$ tensor structure, whose coefficient is given by $2\sqrt{1-\xi}$ in Eq.~(\ref{eq:AtoNLH}). However, it is not clear what ``signal strength" one should use to subtract out the anomalous 4L events, as $\mu_F^{\text{4L}}$ in Table \ref{tab:mucms} is extracted assuming the anomalous coupling vanishes.

Having made these qualifications on our fitting procedures, there are interesting features in Fig.~\ref{fig:xifit} which are likely to survive even after a more rigorous fitting is adopted:
\begin{itemize}
\item The most prominent feature in the right panel of Fig.~\ref{fig:xifit} is that $\xi$ is preferred to be negative, due to the fact that $\mu_F^{\text{4L}}$ in Table \ref{tab:mucms} is larger than 1 in all three different likelihood fits. 
\item In the left panel, the uncertainty in $\xi$ is still rather large, allowing for $\xi \lesssim 0.5$ when ${\cal R}_4^{Z\gamma}$ is turned on or $\xi \gtrsim -0.5$ when ${\cal R}_4^{ZZ}$ is allowed. It turns out that the reason behind the rather loose limits on $\xi$ is different between these two scenarios. 

In the case of ${\cal R}_4^{Z\gamma}$, the loose constraint is due to a combination to two factors: the rather large experimental uncertainty in Table  \ref{tab:Abounds} and a large numerical coefficient, from phase space integration, in front of the $|{\cal R}_4^{Z\gamma}|^2$ term in Eq.~(\ref{eq:R4lAishort}). The large experimental uncertainty implies the differential spectra from $A_4^{Z\gamma}$ are quite similar to those from the leading order $A_1^{ZZ}$. As a consequence, the likelihood fit is unable to separate the two tensor structures. 

On the other hand, while the limit on ${\cal R}_4^{ZZ}$ is the strongest in Table  \ref{tab:Abounds}, implying the likelihood fit is capable of distinguishing this particular tensor structure efficiently, any small presence of $A_4^{ZZ}$ is amplified by the  large numerical coefficient from phase space integration defined in Eq.~(\ref{eq:totw}). This can be seen explicitly either in Eq.~(\ref{eq:R4lAishort}), where the interference term linear in ${\cal R}_4^{ZZ}$ has a numerical coefficient as large as the leading order coefficient.
\end{itemize}

It is possible to further obtain constraints on the Wilson coefficients defined in Eq.~(\ref{eq:efthvv}), by using the mapping in Eq.~(\ref{eq:AtoNLH}). Again we can perform two different fits using the rate and the shape information, respectively. 

\begin{figure}[t]
\begin{center}
\includegraphics[width=4.8cm]{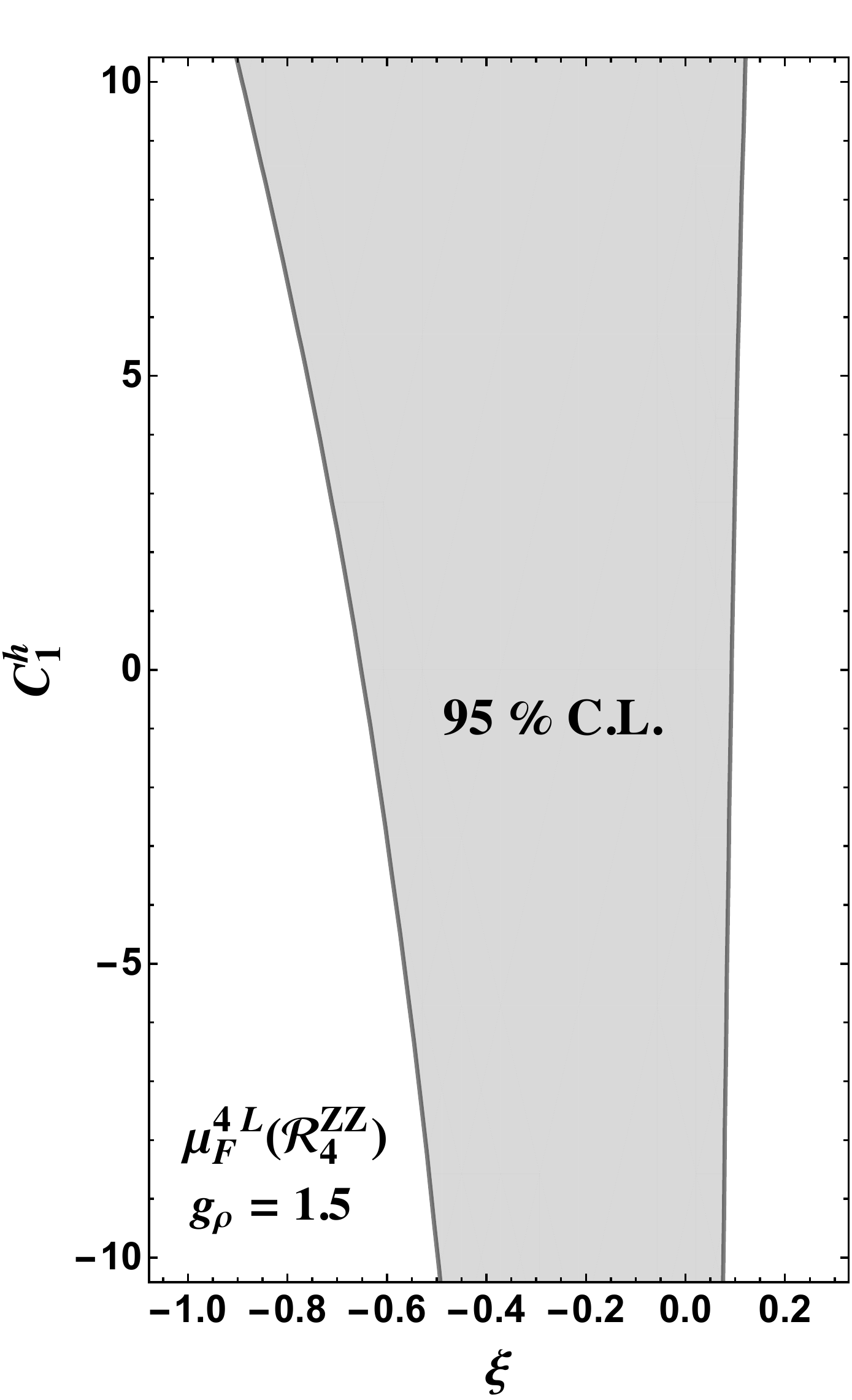}
\includegraphics[width=4.8cm]{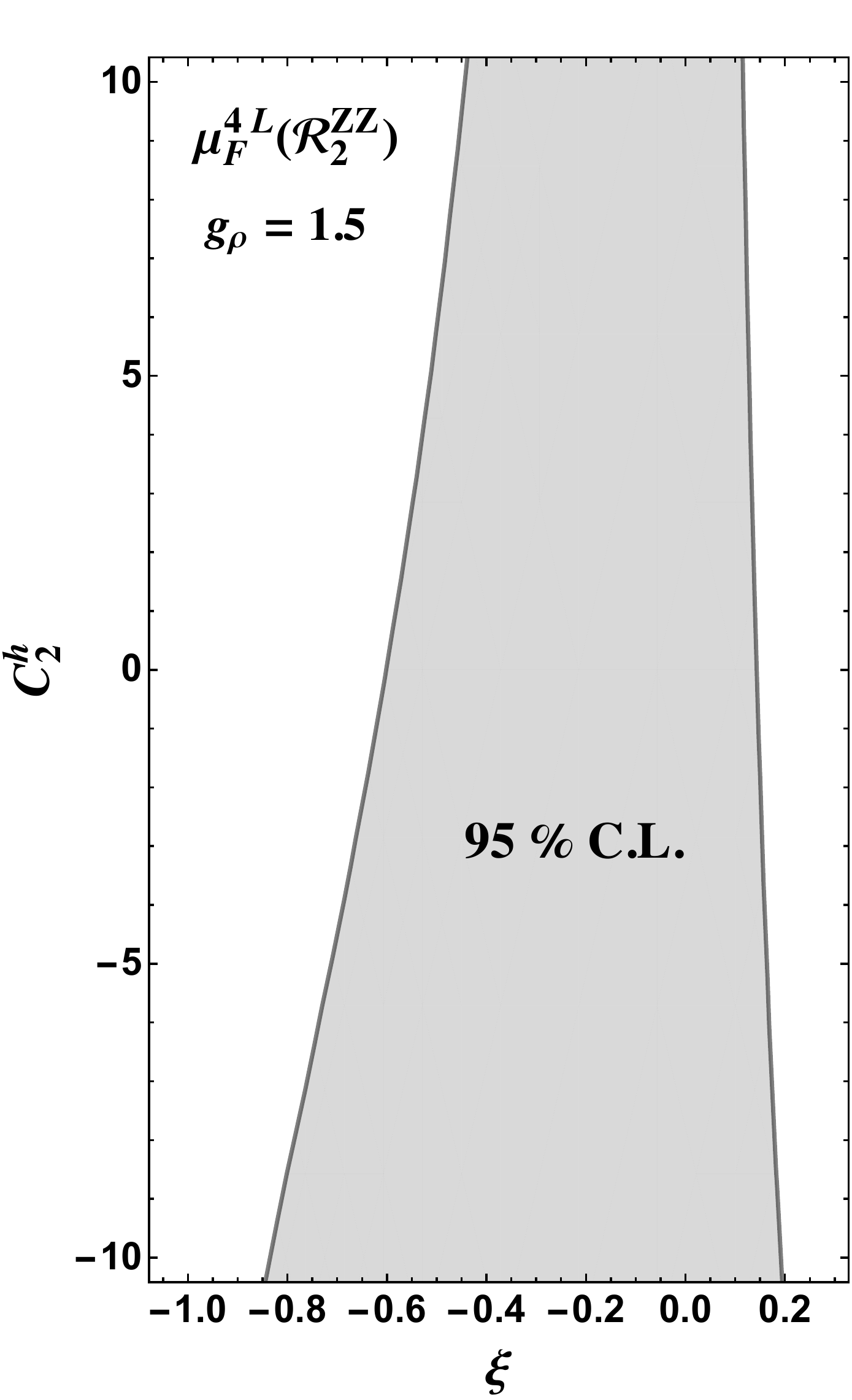}
\includegraphics[width=4.8cm]{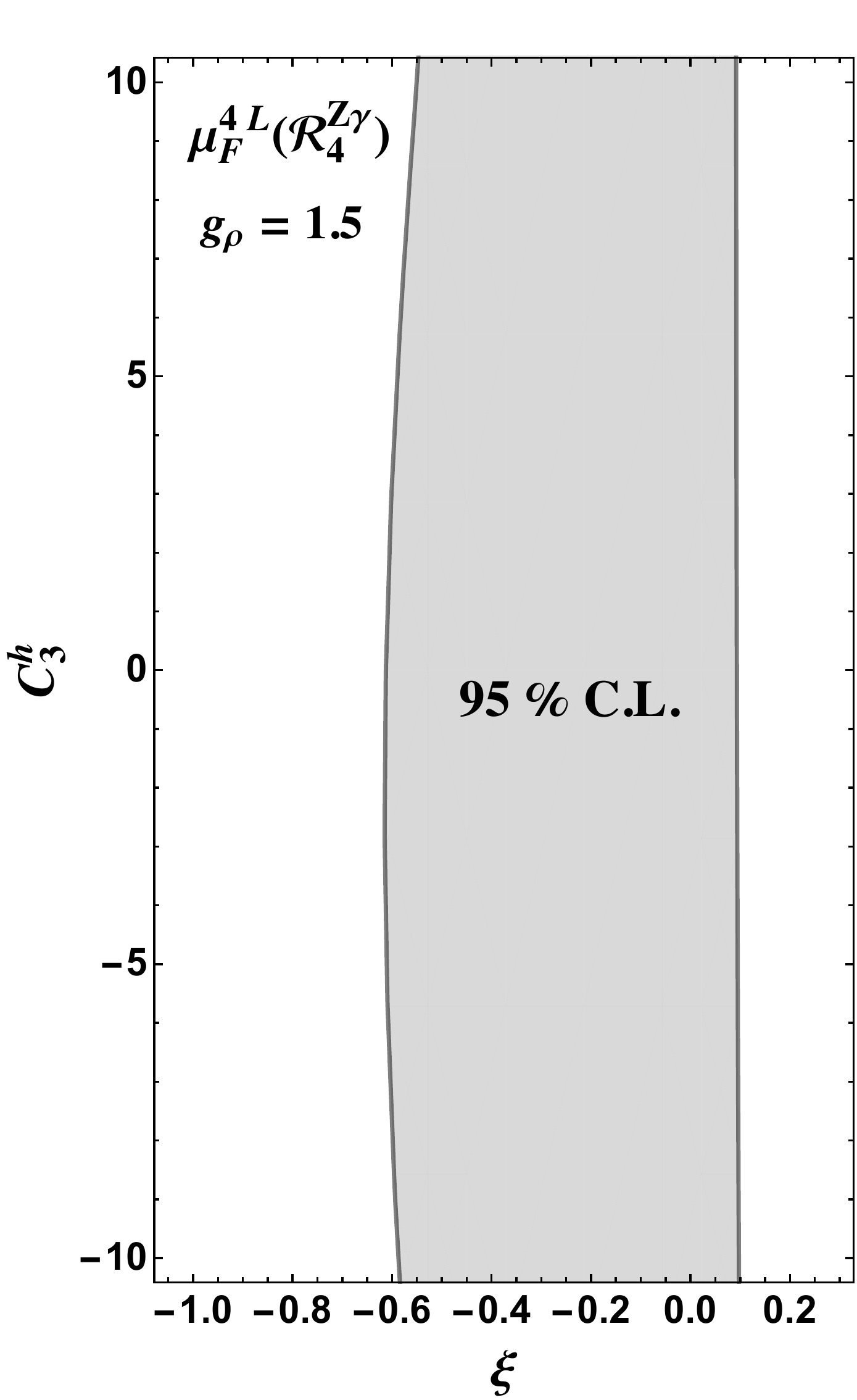}~~\\
~
\includegraphics[width=4.8cm]{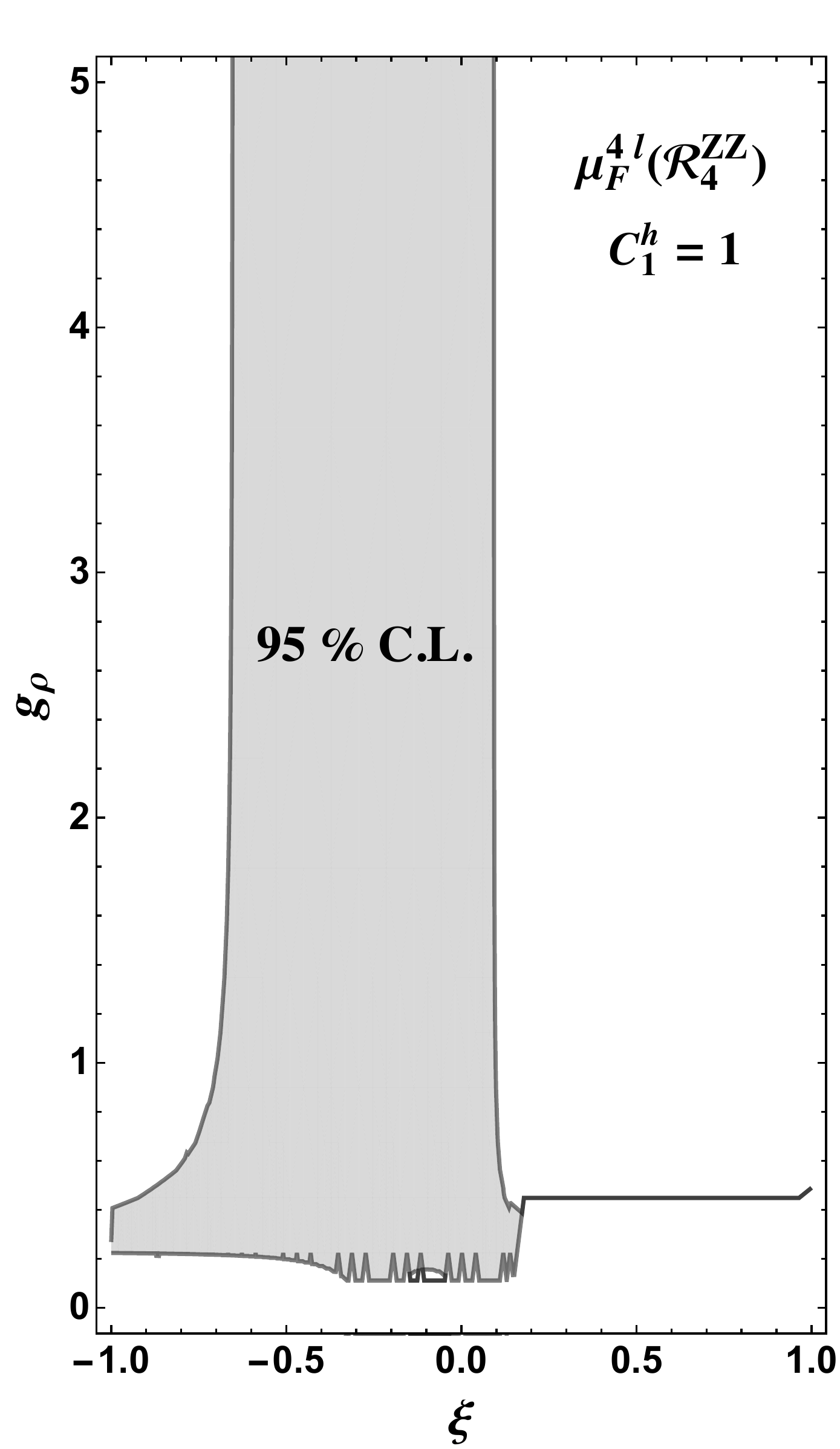}
\includegraphics[width=4.8cm]{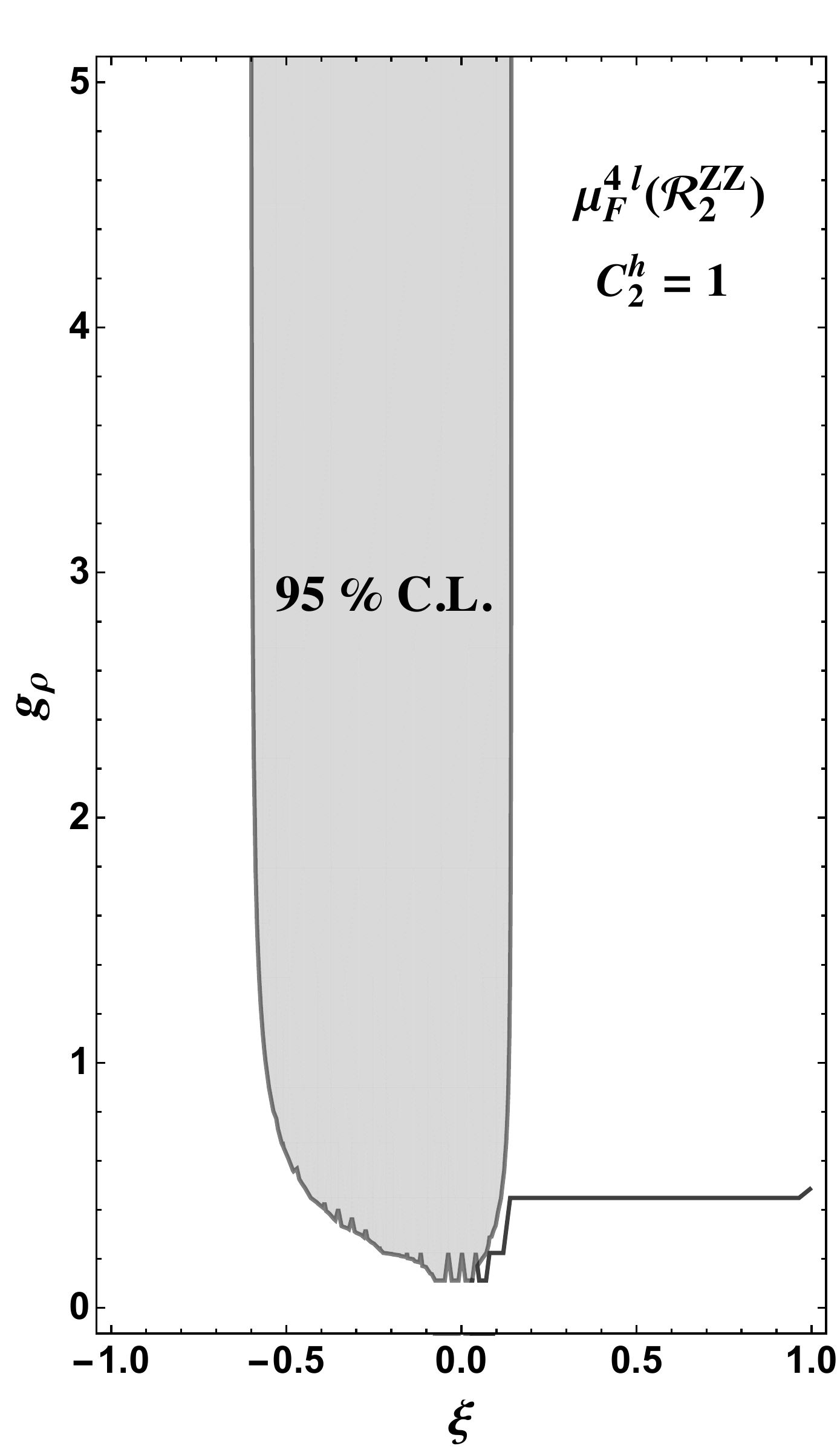}
\includegraphics[width=4.8cm]{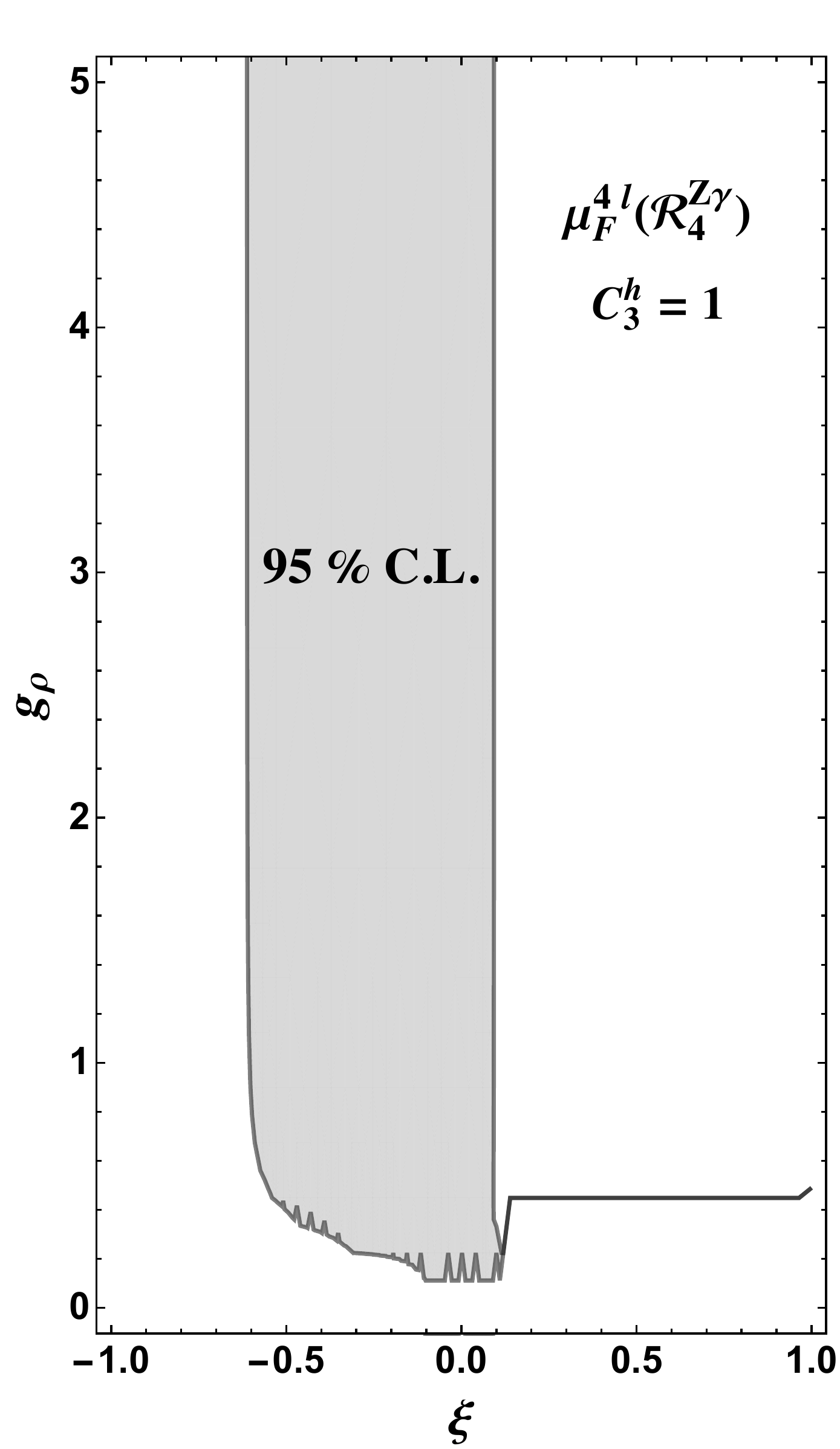}
\caption{\em {\bf Top:}~Allowed parameter space in the $(C_{i}^{h},\xi)$ plane using  using $24.8\ fb^{-1}$ ($7+8$~TeV) + $80.2 fb^{-1}$ (13 TeV) of 4L data \cite{Sirunyan:2019twz}.~{\bf Bottom:}~Same as top, but in the $(g_{\rho},\xi)$ plane.}
\label{fig:CivXi}
\end{center}
\end{figure}

For rate measurements, we re-write Eq.~(\ref{eq:R4lAishort}) in terms of $\xi, g_\rho$ and $C_i^h$,
\bea
\label{eq:R4lCi}
\mu_F^{4L}(C_i)  &=&
(1 - \xi ) 
+  \left[ 0.0519\,  C_1^h
-0.0591\,  C_2^h
-0.00666\,  C_3^h\right] \frac{ \sqrt{1-\xi }\,\xi}{g_{\rho }^2}\nn
&&
+ \left[{0.00278\, (C_1^{h})^2}
+ {0.00456\, (C_2^{h})^2 }
+ {0.00686\, (C_3^{h})^2 } \right.
\nn
&&\left. 
- {0.00313\, C_2^h C_1^h}
- {0.000335\, C_3^h C_1^h}
+{0.000384\, C_2^h C_3^h}\right] \frac{\xi^2}{g_\rho^4}\ ,
\eea
where we have plugged in $M_\rho = g_\rho f$. With this we can use the rate measurements in 4L channel to examine bounds in the $(C_i, \xi)$ and $(g_\rho, \xi)$ two dimensional planes, by fixing the third variable. In the top row of Fig.~\ref{fig:CivXi}, we fix  $g_\rho = 1.5$ and examine the $(C_i, \xi)$ plane. (Recall that in QCD $\alpha_s(m_b)\approx 0.22$, which corresponds to $g_s \approx 1.7$.) We see that turning on $C_i^h$ could have a non-negligible impact on the extraction of $\xi$. However, the impact gets diminished as $g_\rho$ becomes larger and larger because of the $1/g_\rho^2$ and $1/g_\rho^4$ dependence in Eq.~(\ref{eq:R4lCi}). This feature is demonstrated explicitly in the second row of Fig.~\ref{fig:CivXi} where we fix $C_i^h=1$ and plot the constraints in the $(g_\rho, \xi)$ plane.

As for the shape measurements, since the CMS constraints are presented in terms of the percentage of 4L events originating from the anomalous HVV couplings \cite{Khachatryan:2014kca,Sirunyan:2019twz}, we find it convenient to obtain bounds on the ratios $C_i^h/A_1^{ZZ}$, which are independent of the 4L signal strength. The outcome is presented in ~\fref{Cifits}. We see that current bounds are still week and, depending on which coupling is varied in the fit, still allow for large positive or negative values of the $C_i^h$. 
\begin{figure}[t]
\begin{center}
\includegraphics[width=8cm]{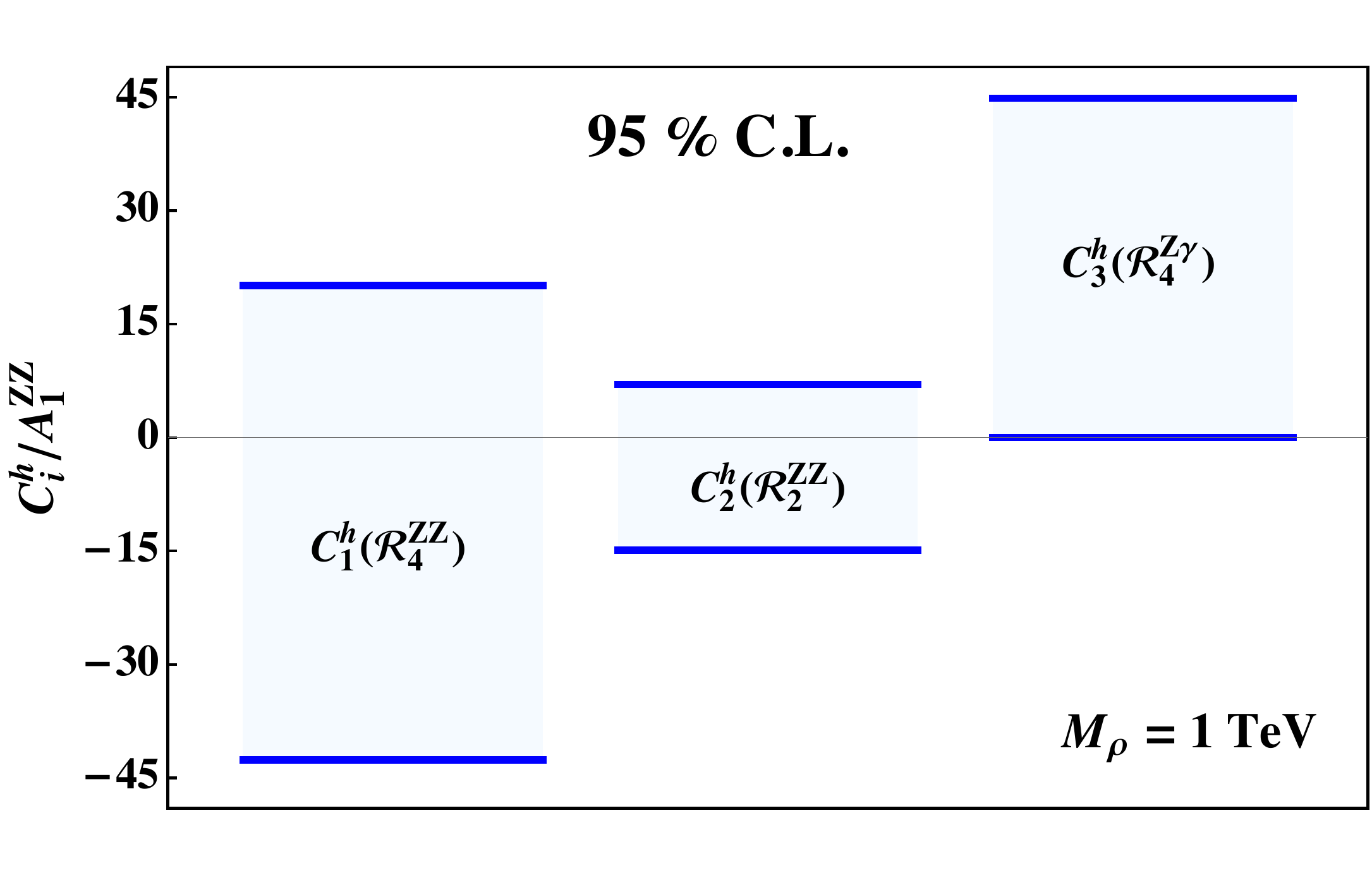}
\caption{\em Fits to $C_i^h/A_1^{ZZ}$ using~\eref{AtoNLH} and taking as input the CMS limits on $\mathcal{A}_i$ shown in~\tref{Abounds}.}
\label{fig:Cifits}
\end{center}
\end{figure}

\subsubsection{Projections with Multi-dimensional Parameter Likelihood Fits}

Currently at the LHC only one parameter fits are performed in the 4L channel \cite{Khachatryan:2014kca,Sirunyan:2019twz}. In the future this channel offers a golden opportunity to conduct multi-parameter fits, as demonstrated in the matrix element method (MEM) framework developed in Refs.~\cite{Chen:2012jy,Chen:2013ejz,Chen:2014pia,Chen:2014gka,Chen:2015iha}. In this framework all decay observables are utilized and a combined likelihood for the $2e2\mu, 4e, 4\mu$ final states is constructed from the \emph{normalized} fully differential decay width. The dominant $q\bar{q} \to$ 4L background, computed analytically in Refs.~\cite{Gainer:2011xz,Chen:2012jy,Chen:2013ejz}, is also included in the likelihood. This likelihood function is a function of all  anomalous HVV couplings in Eq.~(\ref{eq:vert}), and allowing all them to vary simultaneously we can 
 obtain projection curves for future sensitivity to the Wilson coefficients in the nonlinear Higgs Lagrangian by  the mapping in~\eref{AtoNLH}. Details of the statistical analysis and likelihood maximization procedure can be found in~\cite{Chen:2012jy,Chen:2013ejz,Chen:2014pia,Chen:2014gka,Chen:2015iha,Chen:2016ofc}.

In the left panel of \fref{CiMrhocurves} we show projections of the $95\%$\,C.L. contours for $C_i^h/A_1^{ZZ}$ from the shape measurement as a function of luminosity $(\mathcal{L})$, or number of signal events $(N_S)$, at the LHC, assuming $M_\rho=1$ TeV. When calculating the necessary luminosity we include the ggH+VBF production channels at $\sqrt{S}=14$ TeV. The solid lines are projections for an 8-dimensional parameter fit, allowing all anomalous HVV couplings in~\eref{vert} to vary simultaneously while the dashed lines are obtained allowing only a single coupling to vary, as done in current CMS analyses. In the projection we have included $C_4^h$, although it is not currently included in the CMS shape measurements in the 4L channel using Run 2 data. 

On the right-panel we present projections for $M_\rho$ from the shape measurements using  $C_i^h/A_1^{ZZ} = 1/2$. We can see that, in both plots, at large statistics single parameter and multi-parameter fits converge to similar values. However, at low statistics we see that single parameter fits could lead to an overly optimistic sensitivity to the Wilson coefficients and in particular for $C_4^h$. Eventually at the HL LHC~\cite{Cepeda:2019klc}, the sensitivity to each coupling could improve by around an order of magnitude from current limits. As can also be seen, the strongest sensitivity corresponds to $C_4^h$ where values $|C_4^h/A_1^{ZZ}| \sim 4$ can eventually be probed or correspondingly, scales $M_\rho \sim 750$~GeV. This stronger sensitivity is due to the fact the differential spectra corresponding to the $A_2^{Z\gamma}$ can be efficiently distinguished from the dominant tree level $A_1^{ZZ}$ operator~\cite{Chen:2014gka}. This is primarily due to the photon propagator which introduces a pole at low di-lepton invariant mass.~However, in the case of the $A_{4}^{Z\gamma}$ (corresponding to $C_3^h$), the $k_{1,2}^2$ dependence in the tensor structure (see~\eref{vert}) cancels this pole thus eliminating the distinguishing feature in the differential spectra.

\begin{figure}[t]
\begin{center}
\includegraphics[width=8cm]{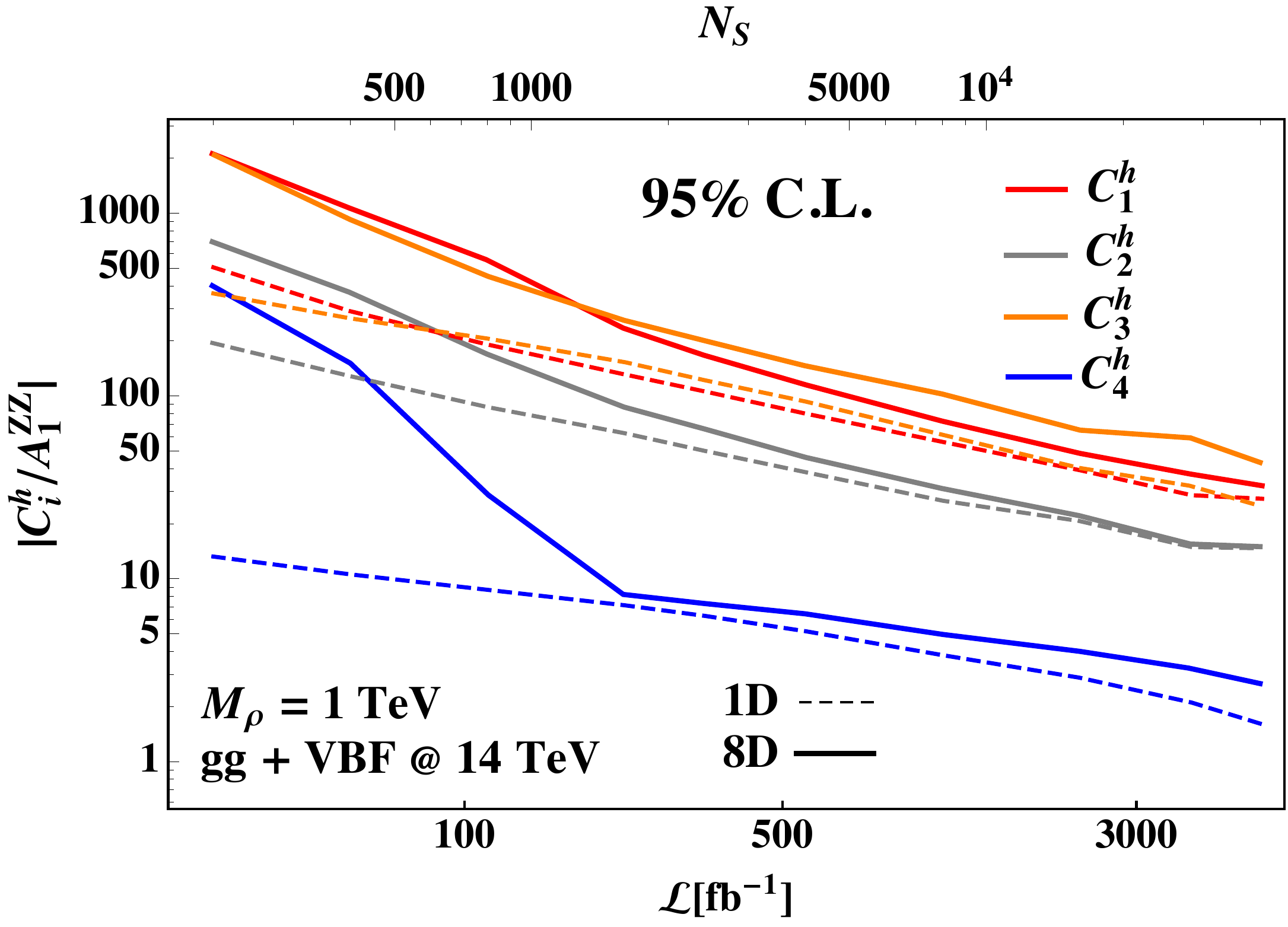}
\includegraphics[width=8cm]{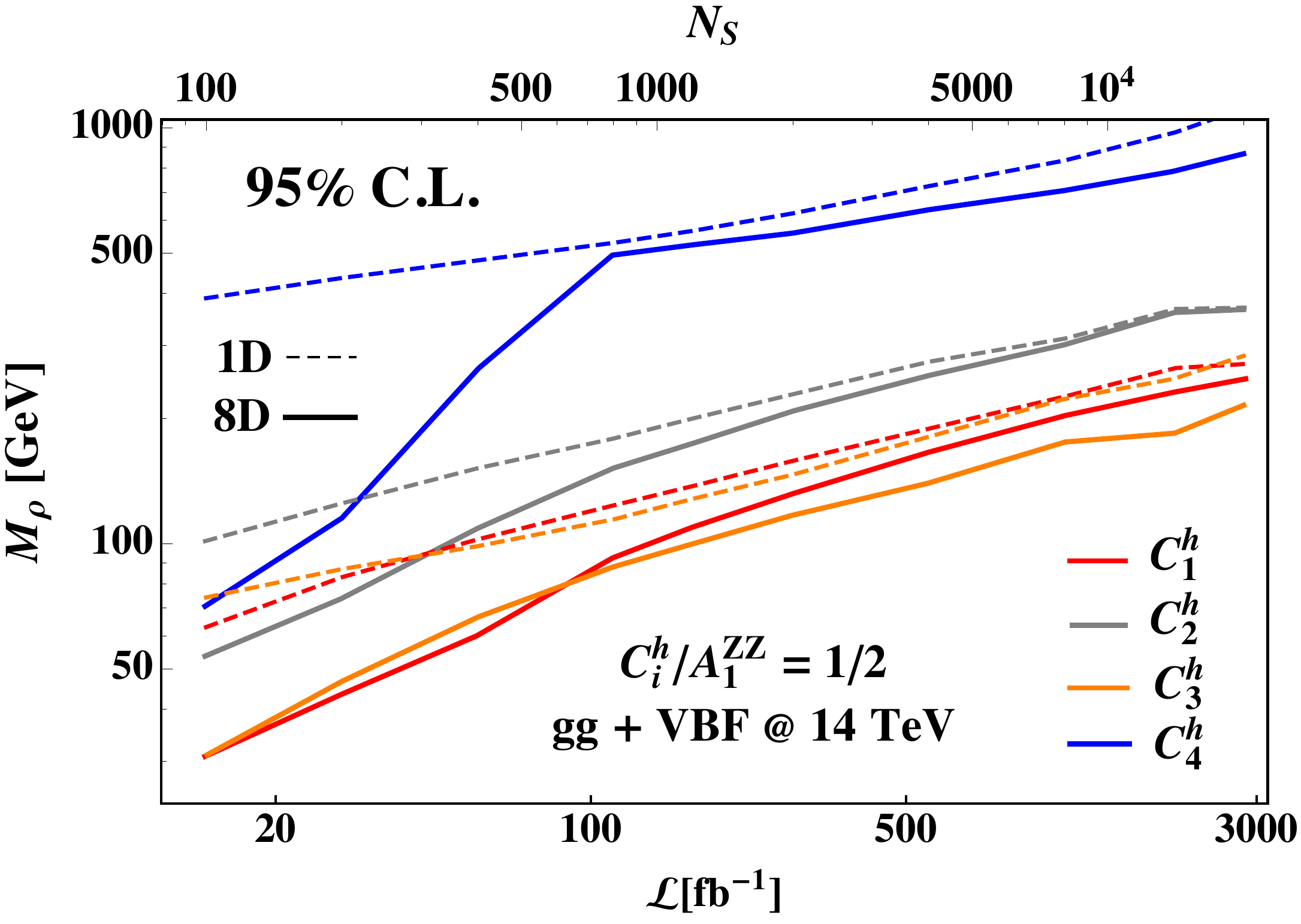}
\caption{\em Projected $95\%$\,C.L. contours for $C_i^h/A_1^{ZZ}$ using shape information  as a function of luminosity or number of signal events $(N_S)$. In the case of luminosity we assume the SM ggH+VBF Higgs production at a 14 TeV LHC and $100\%$ lepton selection efficiency. For these curves the $2e\mu, 4e, 4\mu$ channels are combined and the dominant $q\bar{q}\to4L$ background is also included.}
\label{fig:CiMrhocurves}
\end{center}
\end{figure}

\subsection{Direct H$\,\to\,$Z$\gamma$ Decay}
\label{sect:Zgamma}

\begin{figure}[t]
\centering
      \includegraphics[width=0.5\textwidth]{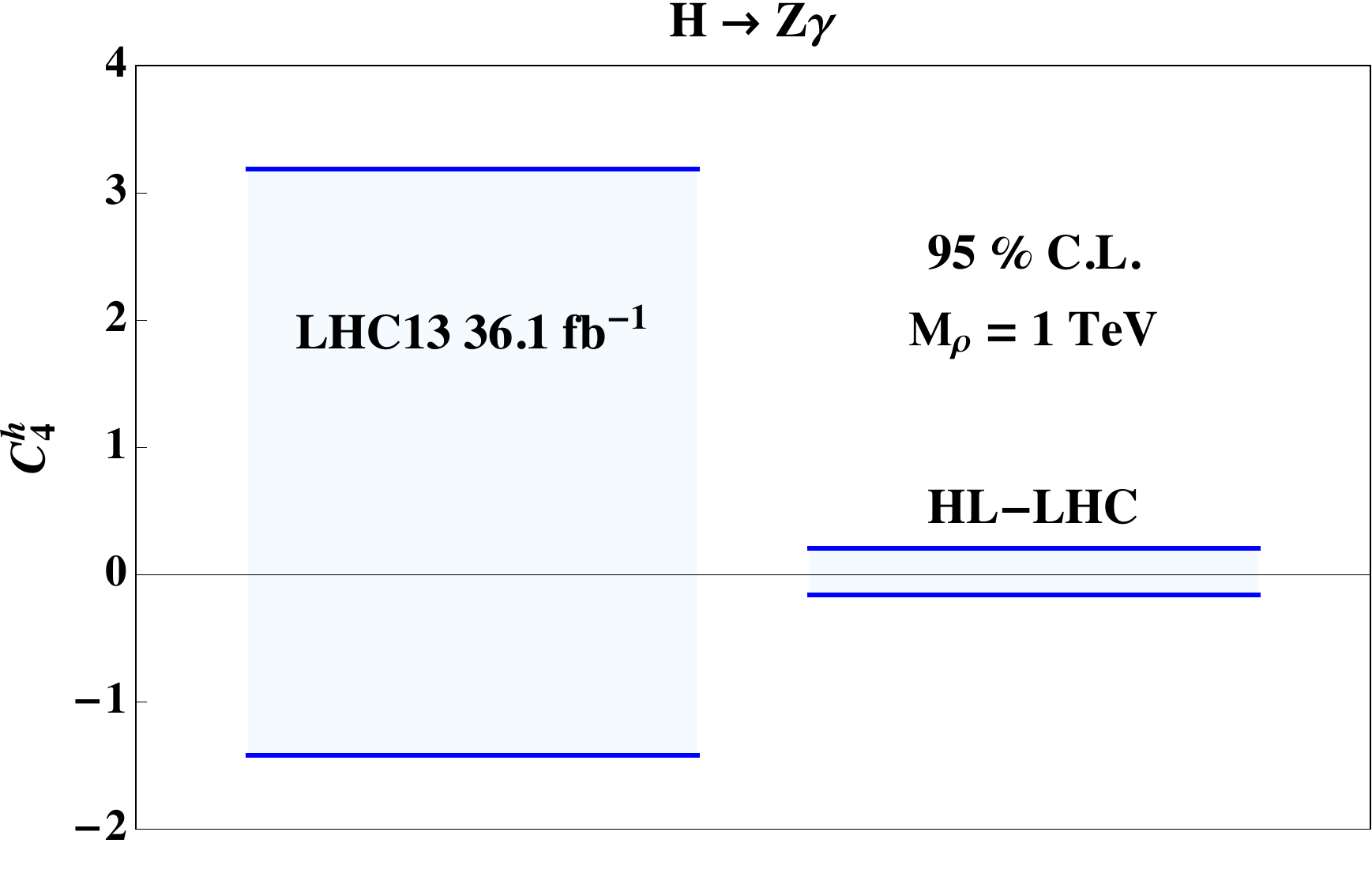}
  \caption{\em The bound on $C_4^h$ from H $\rightarrow$ Z $\gamma$ on-shell measurement at the 13 TeV LHC with luminosity $L = 36.1$ fb$^{-1}$~\cite{Aaboud:2017uhw} and from the prospective HL-LHC~\cite{Dawson:2013bba}.}
  \label{fig:hzg}
\end{figure}

The Wilson coefficient $C_4^h$  contributes to on-shell H $\rightarrow$ Z$\gamma$ decays, which can be parametrized as~\cite{Giudice:2007fh}:
\beq
\mu_{Z\gamma} = \frac{\Gamma(\text{H}\rightarrow \text{Z} \gamma)}{\Gamma(\text{H}\rightarrow \text{Z} \gamma)_{SM}}  \simeq \left|1 - 1.1 \frac{(1\ \  \TeV)^2}{M_\rho^2} C_4^h\right|^2 ,   \\
\eeq
where the  $\mO(1)$ coefficient in front of $C_4^h$ is due to the fact that the SM H $\rightarrow$ Z$\gamma$ process arises at the one-loop level. Although  present measurements  from the LHC  on the signal strength  $\mu_{Z\gamma}$ only gives  $6.6$ at  $95 \%$ CL upper limit \cite{Aaboud:2017uhw,Sirunyan:2018tbk}, it already puts strong constraint on the coefficient $C_4^h$:
\beq
\frac{M_\rho}{\sqrt{|C_4^h|}} > 0.56 (0.84) \ \TeV ,
\eeq
with  $C_4^h > 0 (C_4^h < 0)$ respectively. The HL-LHC will be able to measure this channel with $20 \%$ uncertainty \cite{Dawson:2013bba}, which will in turn give the constraint on $C_4^h$:
\beq
\frac{M_\rho}{\sqrt{|C_4^h|}} > 2.2 (2.5)\ \TeV
\eeq
In Fig.~\ref{fig:hzg}, we  show the $95\%$ C.L. constraint on $C_4^h$ with $M_\rho = $ 1 TeV
from the present measurement with integrated luminosity $L = 36.1$ fb$^{-1}$~\cite{Aaboud:2017uhw} and from the HL-LHC prospective~\cite{Dawson:2013bba}. Note also that when combining the $\mu_{Z\gamma}$ signal strength with the $H\to V^\ast V$ signal strength, their ratio is directly sensitive to the nonlinearity of a composite Higgs boson~\cite{Cao:2018cms}.

\subsection{Precision Electroweak Constraints}
\label{subsec:pew}

While we have demonstrated that precision Higgs measurements, in particular the golden H $\to$ 4L channel, can directly constrain nonlinear Higgs dynamics, it is well-known that there are other low-energy, indirect constraints on anomalous HVV couplings from precision electroweak test (EWPT), which we consider in this section.  The $S,T$ parameters are related to $\xi$ as follows \cite{Barbieri:2007bh}:
\beq
\begin{split}
\label{eq:STIR}
 S_H &=\frac{\xi}{12\pi}\ln\frac{\Lambda^2}{m_h^2}\ ,  \qquad  T_H=-\frac{3}{16\pi}\frac{\xi}{c_w^2}\ln\frac{\Lambda^2}{m_h^2},
 \end{split}
\eeq
where $\Lambda$ is the UV cutoff that regularizes the logarithmic divergence. In the following, we will choose $\Lambda = 4 \pi f$. The fact that the IR contributions to the $S,T$ parameters have different sign put a strong constraint on the value of $\xi$, as the EWPT gives strong positive correlation  $92\%$ among the $S$ and $T$ parameter~\cite{Tanabashi:2018oca}. The present fit to the EWPT gives \cite{Tanabashi:2018oca}:
\beq
S = 0.02 \pm 0.07\ , \qquad T = 0.06 \pm 0.06\ ,
\eeq
which will be used in the following analysis.

In addition to the IR contribution, there are potential UV contributions arising from the presence of $\mO(p^4)$ operators in the effective Lagrangian. In particular,  $O_5^+$ will contribute to the  $S$ parameter at the tree-level. 
\beq
\begin{split}
\label{eq:SToneloop}
 S_{O_5} &=  \frac{32\pi}{g_\rho^2} \xi \ c_5^+(M_\rho)\ .
 \end{split}
\eeq
 There are also contributions from the one-loop threshold corrections involving the vector or fermionic resonances, which are model-dependent~\cite{Contino:2015mha,Contino:2015gdp,Ghosh:2015wiz}. Instead of going into the detail of a particular UV construction,  we simply choose some benchmark values for the these contributions.  Note that the one loop effects can give  negative and positive contributions to the $S$ and $T$ parameters, respectively, which are in the opposite direction from the IR contribution in \Eq{eq:STIR}. This observation can potentially relax the bound on $\xi$ from EWPT.

\begin{figure}[t]
\centering
   \includegraphics[width=0.38\textwidth]{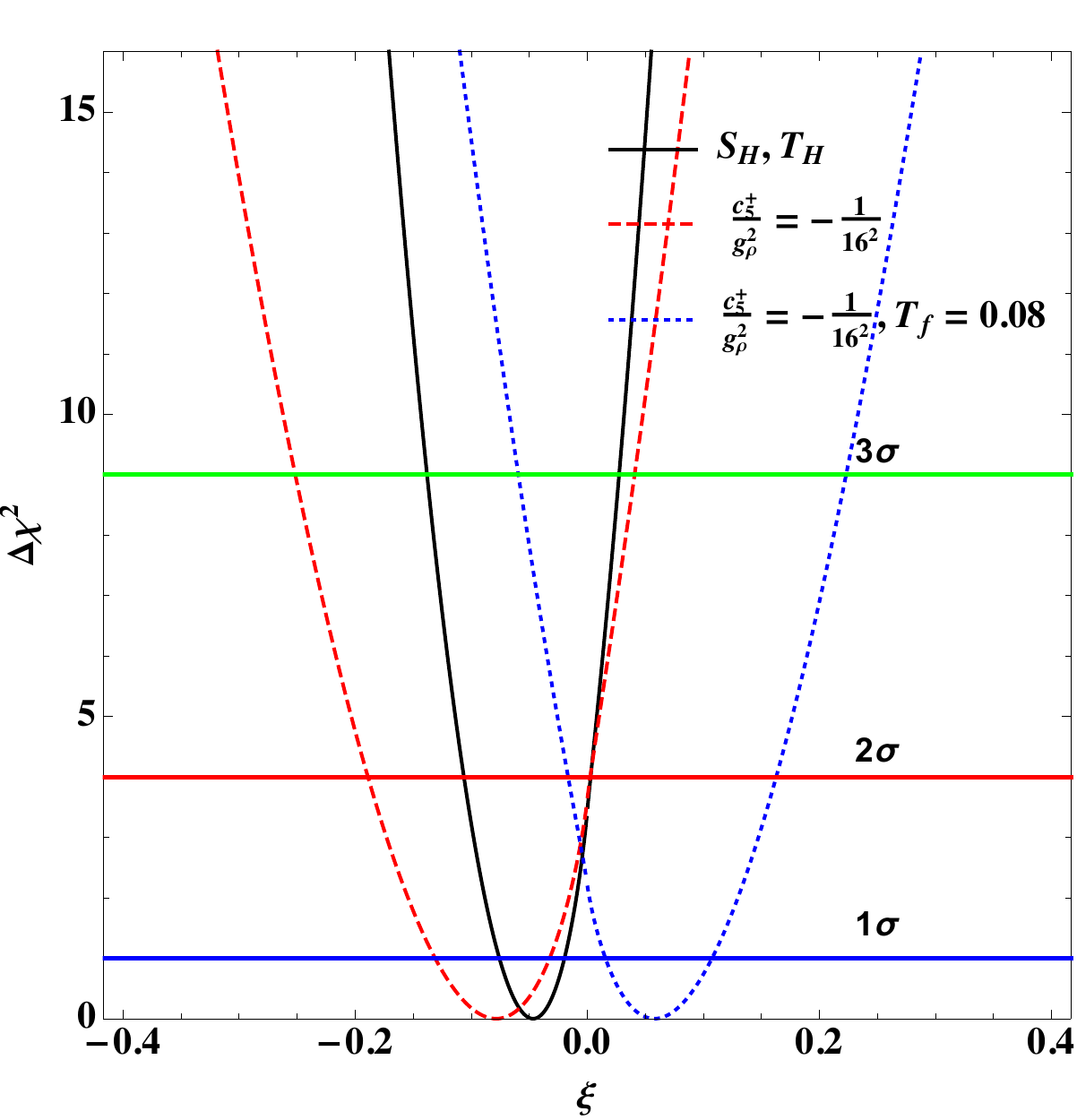}
      \includegraphics[width=0.38\textwidth]{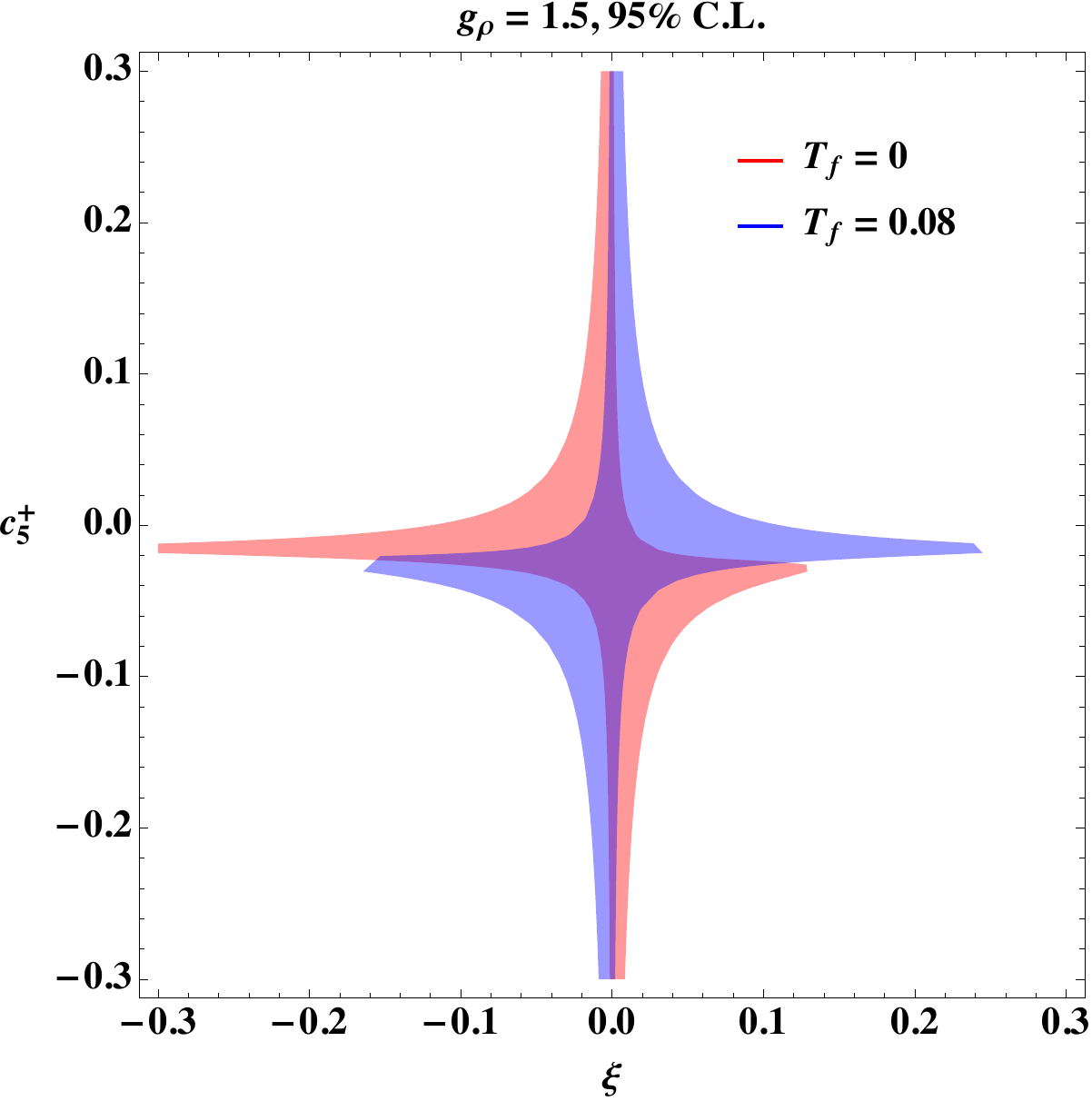}
  \caption{\em Bounds from $S,T$ parameters. Left plot: $\Delta \chi^2$ as a function of $\xi$ under different assumptions in \Eq{eq:assum}. Right plot: $95\%$ C.L. allowed region in the $\xi-c_5^+$ plane with $g_\rho = 1.5$. }
  \label{fig:STbound}
\end{figure}

 In addition, since $\mO_5^+$ contribution to the $S$ parameter arises at the tree-level, as can been seen from the large coefficient in \Eq{eq:SToneloop}, only a small negative $c_5^+$ is needed to relax the bound. They can be achieved by noticing that there are potential cancellations  between different contributions to $c_5^+$ in the Lagrangians of the composite spin-1 resonances \cite{Contino:2011np}:
\beq
c_5^+ = \frac 14 (1 - 4 \, \alpha_2\, g_\rho^2)
\eeq
where $\alpha_2$ is the coefficient of the operator $Q_2$ defined in Ref.~\cite{Contino:2011np}. There are also potential positive contribution to the $T$ parameter coming from the loop of the fermionic resonances, for example, the electroweak singlet top partner~\cite{Lavoura:1992np,Panico:2010is}. We are not going to discuss in detail this possibility, but only take a benchmark value of $T_f = 0.08$ as an illustration. In the left panel of Fig.~\ref{fig:STbound}, we  show the $\Delta \chi^2$ as function of $\xi$ under different three different assumptions: 
\begin{enumerate}
\item $\label{eq:assum} {c_5^+}/{g_\rho^2} = 0, \ T_f = 0$ \ ,
\item ${c_5^+}/{g_\rho^2} = -{1}/{16^2}, \ T_f = 0$\ ,
\item ${c_5^+}/{g_\rho^2} = -{1}/{16^2},  \ T_f = 0.08$.
\end{enumerate}
We can see that the constraint on $\xi$ is very strong in the absence of other contributions: $\xi \in [-0.11,0.002]$ at $95\%$ C.L., which again prefers a negative value of $\xi$. A small negative value of $c_5^+$ relax the bound on the negative $\xi$ to $-0.19$ and additional positive contribution from $T_f$ can relax the bound on positive value of $\xi \in [-0.015,0.16]$. In the right panel we present the $95\% $ C.L. allowed region in the $\xi - c_5^+$  plane with $g_\rho = 1.5$. We can see that the bound on $\xi$ is very strong with only IR contribution and can be significantly relaxed in the presence of small negative value of $c_5^+$ and positive contribution $T_f$, which is consistent with the results discussed above.

We emphasize that the precision electroweak constraints are orthogonal to direct measurements of Higgs couplings, as they involve different sets of assumptions. Therefore it is important to pursue both of them independently.

\section{Conclusions}\label{sec:conc}

Nonlinear dynamics of the 125 GeV Higgs boson is the most salient feature of a composite Higgs boson. The nonlinear interaction realize the PNGB nature of the Higgs, in the same way pions in low-energy QCD manifest their PNGB nature through the nonlinear interactions. In fact, the nonlinear dynamics can serve as the defining property of a composite Higgs boson, independent of how the fermionic sector is implemented, even when the so-called top partner is neutral under all SM charges. In particular, recent theoretical advances showed the nonlinear interaction is an IR property of the composite Higgs boson that is insensitive to the symmetry-breaking pattern invoked in the UV. At the two-derivative level, the nonlinear Lagrangian is determined by a single parameter $f$, the Goldstone decay constant. In the low-energy regime, $f$ needs to be taken as an input from the experimental data. Traditionally this is done by relating $f$ to the signal strength of HVV couplings assuming the observed events arise entirely from ${\cal O}(p^2)$ operators in the nonlinear Higgs Lagrangian. This is obviously an over-simplification as ${\cal O}(p^4)$ operators introduce several new anomalous tensor structures to HVV couplings.

In order to disentangle effects of ${\cal O}(p^2)$ operators from those of ${\cal O}(p^4)$ operators, it is crucial to include shape information, as different tensor structures lead to different shapes in the fully differential spectra. In this regard, the H$\to$ 4L ``golden channel'' decay is the ideal probe of these nonlinear Higgs dynamics. In this work we have performed, for the first time, experimental fits to the Wilson coefficients of ${\cal O}(p^4)$ operators. In addition, we demonstrated the limitation of using only the signal strength to constrain the nonlinear parameter $\xi$ at leading order in derivative expansions. We showed that it is important to include both the rate information in the signal strength measurements and the shape information in the fully differential spectra, in order to constrain the nonlinear parameter $\xi$ and the Wilson coefficients $C_i^h$. The fitting procedures we adopted are less than ideal and limited by the methods employed in current LHC analyses.

We found that in rate measurements $\xi$ is preferred to be negative, pointing to a non-compact coset structure in the UV  \cite{Low:2014nga,Low:2014oga}. Moreover, using the shape measurements we identified scenarios where $\xi$ could be as large as +0.5 or -0.5, corresponding to turning on $A_4^{Z\gamma}$ and $A_4^{ZZ}$, respectively. Such a large $\xi$ could be further constrained in a specific UV model, either by precision electroweak measurements or direct searches in the fermionic sector. It would be an interesting model-building challenge to devise a concrete UV model realizing such a large $\xi$ while avoiding other experimental constraints.

Our work points to a new experimental frontier in using precision Higgs measurements to probe nonlinear dynamics of a composite Higgs boson, beyond the conventional signal strength measurements. Given that the Higgs boson remains a top priority in current and future experimental programs in high-energy colliders, it is desirable to introduce new tools and techniques to further the experimental analysis in measurements of Higgs properties. We leave this direction for future works.

\begin{acknowledgments}
We are grateful to Yi Chen for his help in obtaining some of the results presented in this work and Daniel Stolarski for useful comments. This work is supported in part by the U.S. Department of Energy under contracts No.~DE-AC02-06CH11357 and No.~DE-SC0010143. The work of R.V.M.~is supported by MINECO grants~FPA 2016-78220-C3-1-P,~FPA 2013-47836-C3-2/3-P (including ERDF), and the Juan de la Cierva program,~as well as by the Junta de Andalucia Project FQM-101.

\end{acknowledgments}


\bibliography{NonLinearHiggsRefs}

\end{document}